\documentclass[10pt,aps,pre,twocolumn,superscriptaddress,floatfix,longbibliography]{revtex4-1}
\usepackage{graphicx}
\usepackage{amsmath}
\usepackage{amssymb}
\usepackage{hyperref}
\begin{document}

\title[Sedimentation of a suspension of rods]{Sedimentation of a suspension of rods: Monte Carlo simulation of a continuous two-dimensional problem}

\author{Nikolai I. Lebovka}
\email[Corresponding author: ]{lebovka@gmail.com}
\affiliation{Department of Physical Chemistry of Disperse Minerals, F. D. Ovcharenko Institute of Biocolloidal Chemistry, NAS of Ukraine, Kiev, Ukraine, 03142}
\affiliation{Department of Physics, Taras Shevchenko Kiev National University, Kiev, Ukraine, 01033}
\author{Yuri Yu. Tarasevich}
\email[Corresponding author: ]{tarasevich@asu.edu.ru}
\affiliation{Laboratory of Mathematical Modeling, Astrakhan State University, Astrakhan, Russia, 414056}
\author{Leonid A. Bulavin}
\affiliation{Department of Physics, Taras Shevchenko Kiev National University, Kiev, Ukraine, 01033}
\author{Valery I. Kovalchuk}
\affiliation{Department of Physics, Taras Shevchenko Kiev National University, Kiev, Ukraine, 01033}
\author{Nikolai V. Vygornitskii}
\affiliation{Department of Physical Chemistry of Disperse Minerals, F. D. Ovcharenko Institute of Biocolloidal Chemistry, NAS of Ukraine, Kiev, Ukraine, 03142}
\date{\today}

\begin{abstract}
The sedimentation of a two dimensional suspension containing rods was studied by means of
Monte Carlo (MC) simulations. An off-lattice model with continuous positional and orientational
degrees of freedom was considered. The initial state before sedimentation was produced using
a model of random sequential adsorption. During such sedimentation, the rods undergo translational
and rotational Brownian motions. The MC simulations were run at different initial number densities
(the numbers of rods per unit area), $\rho_i$, and sedimentation rates, $u$. For sediment films, the spatial
distributions of the rods, the order parameters and the electrical conductivities were examined. Different types of sedimentation-driven self-assembly and anisotropy of the electrical conductivity were revealed inside the sediment films. This anisotropy can be finely regulated by changes in the values of $\rho_i$ and $u$.
\end{abstract}

\maketitle

\section{Introduction\label{sec:intro}}

Gravitational or centrifugal techniques are widely used in practice for the separation of suspended particles~\cite{Schaflinger1990}, industrial filtration~\cite{Tiller1995}, the clarification of solvents and the flotation of suspended sewage solids~\cite{Chen1997}, the production of colloidal phases~\cite{Chen2015}, and the preparation of electrically conductive and transparent films~\cite{Petek2015}. An important characteristic of the deposition process is the sedimentation length, $\lambda$, as determined by Perrin in measurements of the Boltzmann constant, $k_B$~\cite{Perrin1913}. This value is defined~as
\begin{equation}\label{eq:lambda}
  \lambda = \frac{k_BT}{mg},
\end{equation}
where $k_BT$ is the thermal energy, $m$ is the buoyant mass of the particle and $g$ is the acceleration. The buoyant mass of the particle is defined as $m = \Delta\rho V$, where $\Delta\rho$ is the difference between the densities of the particle and the liquid, and $V$ is the volume of the particle. In very dilute systems (ideal gaslike systems), the equilibrium sedimentation profile of particle number density, $\rho$,  is barometric and decays exponentially with height, $\rho=\rho_0 \exp(-y/\lambda)$, where $\rho_0$ is the density at the base of sediment, $\rho_0 = \rho(0)$. This approximation works when the value of $\lambda$ is large compared to the inter-particle correlation length, but small compared to the vertical size of the system.

For an isolated sphere of radius $r$ with no-slip boundary conditions on the surface, the sedimentation rate can be evaluated using Stokes law
\begin{equation}\label{eq:u}
u = \frac{mg}{f_t},
\end{equation}
where $f_t = 6\pi\eta r$ is the Stokes' friction coefficient related to the translation diffusion coefficient~as
\begin{equation}\label{eq:Dt}
D_t = \frac{k_BT}{f_t},
\end{equation}
and $\eta$ is the viscosity of surrounding liquid~\cite{Schuck2016}.

The P\'{e}clet number is defined as the ratio of the Brownian, $\tau_B$, and sedimentation, $\tau_S$, times of a particle over the same distance (e.g., a distance of one radius), i.e. $\mathrm{Pe}=\tau_B/\tau_S$~\cite{Royall2007}. Using the values $\tau_B=r^2/D_t$ and $\tau_S=r/u$, we can obtain $\mathrm{Pe}= r u / D_t$. Accounting Eqs.~\eqref{eq:lambda}, \eqref{eq:u}, and \eqref{eq:Dt},
\begin{equation}\label{eq:Pe}
\mathrm{Pe}= \frac{r}{\lambda},
\end{equation}
i.e., the P\'{e}clet number simply presents the ratio of particle size $r$ and sedimentation length $\lambda$.

Uniformity in particle arrangement is expected when the effects of diffusion are dominant ($\mathrm{Pe}\ll 1$), whereas in the opposite case, when the effects of sedimentation are dominant ($\mathrm{Pe}\gg 1$) a spatial gradient in the distribution of particles is typically observed. For example, at room temperature, $T = 298$ K, $\Delta\rho= 10^3$ kg/m$^3$, and $g = g_0 = 9.8$ m/s$^2$ ($g_0$ is the acceleration due to gravity) at $r = 0.5$ $\mu$m, we have $\lambda\approx 0.8$ $\mu$m ($\mathrm{Pe}\approx 0.62$), i.e., a uniform arrangement, while, at $r = 15$ $\mu$m, we have $\lambda\approx 0.1$ $\mu$m ($\mathrm{Pe}\approx 10$), i.e., this represents a non-uniform arrangement. The experimental investigations performed for regimes of $\mathrm{Pe}\ll 1$~\cite{Thies-Weesie1995} and $\mathrm{Pe}\gg 1$~\cite{Segre2001} revealed different types of particle arrangement. Computer simulation methods in different P\'{e}clet regimes have also been applied to study the steady state sedimentation of particles with spherical~\cite{Royall2007,Padding2004} and non-spherical~\cite{Wachs2009,Choi2013,Karimnejad2018,Singh2018} shapes.

There is special interest in the sedimentation effects in suspensions of impenetrable rodlike particles (rods). For example, electrically conductive and transparent films prepared from sediments of carbon nanotubes are of particular interest in the production of electrodes for super-capacitors, thin film transistors, and fuel cells~\cite{Hu2010}.

For systems of elongated particles, self-assembly and phase transition effects can introduce supplementary complications to the sedimentation regimes. For example, experimental studies have revealed the presence of jamming and orientational ordering in colloidal rod sedimentation~\cite{Mohraz2005}. A density-driven isotropic-nematic (IN) phase transition in three dimensional (3D) homogeneous system of rods with infinite aspect ratio (length-to-diameter ratio $a=\infty$) was theoretically predicted in the 1940s~\cite{Onsager1949}. The theory predicted coexisting isotropic and nematic phases between particle number densities $\rho_i\approx 3.29$ and $\rho_n\approx 4.19$, and a transition to the nematic phase with strong ordering at $\rho\geq \rho_n$. Spontaneous ordering in concentrated rod suspensions has been experimentally confirmed in many studies~\cite{Adams1998,Maeda2003,Alargova2004}.

In two-dimensions (2D), the Onsager theory predicts a continuous IN transition at a critical density of $\rho_i=3\pi/2\approx 4.7$~\cite{Kayser1978}. Monte Carlo (MC) simulations have revealed the IN transition as the number density of rods, $\rho$, increases~\cite{Frenkel1985,Bates2000}. The ordered phase becomes absolutely unstable with respect to disclination unbinding at $\rho<\rho_i$, but a transition to a quasi-nematic phase at $\rho\geq\rho_n \approx 7.25$ occurs. The quasi-nematic phase demonstrates algebraic order (quasi long-range order) and the occurrence of a disclination-unbinding transition of the Kosterlitz--Thouless (KT) type has been suggested. MC simulations of continuous 2D systems of rods revealed a 2D nematic phase of the KT type at high densities for relatively long rods with the high aspect ratio of  $a\gtrsim 7$~\cite{Bates2000}. For non-equilibrium 2D systems of rods obtained using a random sequential adsorption (RSA) model, further self-assembly is possible owing to deposition-evaporation processes or to the diffusion motion of particles. Several problems related to such types of self-assembly of rods have previously been discussed~\cite{Ghosh2007,Lopez2010,Loncarevic2010,Matoz-Fernandez2012,Kundu2013,Kundu2013a,Lebovka2017}.

Pioneering work on computer simulation of the sedimentation of dilute dispersions of randomly oriented rods was performed in 1959~\cite{Vold1959}. It was demonstrated that the volume fraction of rods in the sediment decreases significantly with increase of the aspect ratio, $a$, of the particles. Self-assembly can be important at late stages of the sedimentation when high density layers are formed at the bottom. For example, pronounced density oscillations near the base have been observed in system of rods~\cite{Biben1993}.

Sedimentation of non-Brownian rods complicated by the long-range nature of multibody hydrodynamic interactions has intensively been studied in theoretical, experimental, and simulation works (for a review, see~\cite{Guazzelli2011}). The mechanism of instability and cluster formation in suspension of non-spherical sedimenting particles was described in~\cite{Koch1989}. Theoretical calculations predicted that the hydrodynamic interactions could result in coupling between the particles orientations. Therefore, the flow field generated by the particles can lead to a clustering of the particles and enhancement of the sedimentation rate.

The experimental investigation of the sedimentation of a dilute non-Brownian fibers within a viscous fluid (at very low Reynolds number, $\mathrm{Re}\rightarrow 0$) have been performed under well stirred conditions~\cite{Herzhaft1996}. The aspect ratio of fibers was $a\approx10$ and effective number density was $\rho=0.09(2/l)^3$ (here $l$ is the length of the particle). The well-stirred suspensions were unstable, formation of particle clusters was observed; within a cluster, individual particles tend to align in the direction of gravity. These investigations were continued for the fibers with different aspect ratios ($a=5$--32) and different concentrations of suspensions ($=(0.019-13)\times(2/l)^3$)~\cite{Herzhaft1999}. The different regimes of sedimentation were identified. The velocity fluctuations were found to be large and anisotropic. Formation of large-scale streamers from clusters of fibers during early stages of the sedimentation was experimentally observed~\cite{Metzger2005,Metzger2007}. The streamers presented highly correlated regions of downward velocities. The `clumps' of parallel non-Brownian fibers settled more rapidly than individual fibers.

The MC~\cite{Mackaplow1998}, dynamics simulations~\cite{Butler2002,Saintillan2005}, and theoretical kinetic model~\cite{Helzel2017} also predicted the formation of clusters and migration of fibers into narrow streamers aligned in the direction of gravity at $\mathrm{Re}\rightarrow 0$. The 3D dynamical simulations has been performed to study the sedimentation of large prolate non-Brownian spheroids of aspect ratio $a = 5$ at small, but nonzero Reynolds number ($\mathrm{Re}=0.3$)~\cite{Kuusela2001,Kuusela2003}. The results revealed that, in the diluted suspension, the inertial effects tend to align particles horizontally (i.e., perpendicular to gravity), whereas, with increasing concentration, the orientation of particles becomes opposite (i.e., parallel to gravity). These sedimentation data were qualitatively supported by the experimental investigations of sedimentation of non-Brownian fibers at $\mathrm{Re}\rightarrow 1$~\cite{Salmela2007}.

A computational hydrodynamic 2D model was applied to simulate the suspensions of rigid thin fibers in a viscous incompressible fluid with a nonzero Reynolds number and in absence of Brownian motion~\cite{Wang2009}. The data confirmed the fiber clustering in the small Reynolds number regime; by contrast, the dispersion instead of aggregation was observed in the high Reynolds number regime. The large scale 3D numerical simulations of the gravity induced sedimentation of non-Brownian fibers in a highly viscous fluid revealed existence of the {densification} phase (where the cluster densifies and grows), and the \emph{coarsening} phase (where the cluster becomes smaller and less dense)~\cite{Gustavsson2009}.

The Brownian motion randomizes the orientational ordering of fibers, however, at large concentration, the Onsager type of nematic ordering can also be important due to the entropic reasons~\cite{Vroege1992,Philipse1997}. The Onsager theory and its Parsons--Lee extension have been applied for calculation of the concentration profiles of rods with aspect ratios over the range $a = 10$--80 in a gravitational field~\cite{Allen1999}. The calculations revealed nematic ordering at the bottom of the sediment and a strong dependence of the calculated profiles on the value of $a$. The MC simulations were applied to investigate sedimentation and phase equilibria in suspensions of hard spherocylinders with aspect ratio of~\cite{Savenko2004}. The investigations revealed that sedimentation led to multi-phase coexistence of isotropic and different liquid crystalline phases. However, previous studies of sedimentation of Brownian rods have not paid much attention to the effects of the particle density and sedimentation rate on the kinetics of sedimentation, on evolution orientational ordering and the electrical conductivity of the sediments.

This paper analyzes the self-assembly of Brownian rods during sedimentation in 2D colloidal suspensions. The MC simulations have been applied. Note that the MC technique is not aimed at generating dynamics. However, in the limit of long times, MC produces the trajectories equivalent to trajectories of a system under Brownian dynamics. Generally the standard MC is more fast (about one order of magnitude) than the Brownian dynamics~\cite{Saintillan2005}. Moreover, the proper rescaling the MC accounting for the acceptance rate of simultaneous trial displacements and rotations allows a direct comparison between MC and Brownian dynamics simulations~\cite{Sanz2010,Romano2011,Patti2012,Cuetos2015}. Recently such approach have been applied to study the rodlike particles in the isotropic phase~\cite{Corbett2018}.
The initial state was produced using an RSA model with isotropic orientations of the rods, after which the sedimentation was started and the rods were allowed to undergo both translational and  rotational diffusion. The kinetics of the changes of structure and the electrical conductivity in the horizontal $x$ and vertical $y$ directions have been analyzed.

The rest of the paper is constructed as follows. In Sec.~\ref{sec:methods}, the technical details of the simulations are described, all necessary quantities are defined, and some test results are given. Section~\ref{sec:results} presents our principal findings. Section~\ref{sec:conclusion} summarizes the main results.

\section{Computational model\label{sec:methods}}

The initial state before sedimentation was produced using an RSA model~\cite{Evans1993}. Rods of length $l$ and thickness of $d$ with a large aspect ratio, $a=l/d \gg 1$ were deposited onto a plane randomly (both their positions and orientations were random) and sequentially until the desired initial number density $\rho_i$ (i.e., number of rods per unit area) were obtained. Their overlapping with previously placed particles was forbidden. An isotropic initial orientation of the rods was assumed. Since the aspect ratio is large, we ignored the finite width of the rods when looking for the rod intersections, i.e., in this case, the rods were treated as zero-width objects.

The length of the system was $L$ along the horizontal $x$-direction and periodic boundary conditions were applied along this axis. The height of the system along the vertical $y$-direction was $H$. In the present work, all calculations were performed using $L=H=32 l$. Zero flux boundary conditions were applied at the upper and lower borders.

The simulation of the sedimentation assumed simultaneous Brownian motion of the rods and their downward movements. The Brownian diffusion of rods was simulated using the MC procedure. At each step, an arbitrary rod was randomly chosen, and its rotational and translational diffusion motions were taken into consideration.  The rotational diffusion coefficient was calculated as $D_r=k_BT/f_r$, the translational diffusion coefficients were calculated as $D_\parallel=k_BT/f_\parallel$ and $D_\perp=k_BT/f_\perp$ (compare with Eq.~\eqref{eq:Dt} for spherical particles) for the motions along and perpendicular to the direction of the long axis, respectively.
Here $f_r$, $f_\parallel$, and $f_\perp$  are the Stokes' friction coefficients for rotational and translational motions. For long rods, $a\gg 1$,  the following formulas can be used~\cite{Loewen1994}
\begin{subequations}\label{eq:f}
\begin{align}
f_r &= \frac{\pi\eta l^3}{3(\ln a+\gamma_r)},\\
f_\parallel &= \frac{2\pi\eta l}{\ln a+\gamma_\parallel},\\
f_\perp &= \frac{4\pi\eta l}{\ln a+\gamma_\perp},
\end{align}
\end{subequations}
where $\eta$ is the viscosity of the surrounding liquid, and
$\gamma_r\approx -0.662$,
$\gamma_\parallel \approx -0.207$, and
$\gamma_\perp \approx 0.839$ are the end correction coefficients.

The amplitudes of the Brownian motions are proportional to the square root of the corresponding diffusion coefficients. The amplitude of the displacement along the long axis was put as $\Delta r_\parallel= \alpha l$, where  $\alpha$ was chosen to be small enough ($\alpha=0.05$) in order to obtain satisfactory acceptance of the MC displacement~\cite{Landau2014}.

The other amplitudes were evaluated using the following equations
\begin{subequations}\label{eq:rt}
\begin{align}
\Delta r_\perp  / \Delta r_\parallel &= \sqrt{f_\parallel/f_\perp},\\
l \Delta \theta  / \Delta r_\parallel &= \sqrt{f_\parallel/f_r}.
\end{align}
\end{subequations}
For example, at $a=10^3$, we have (see Eqs.~\eqref{eq:f}--\eqref{eq:rt})  $\Delta r_\perp/\Delta r_\parallel= 0.76 $ and  $l \Delta \theta /\Delta r_\parallel= 2.37\alpha$.

One MC time step ($\Delta t_{MC}=1$) corresponds to the attempted two displacements and one rotation for the all rods in the system.  This time increment corresponds to the Brownian dynamics time increment that can be evaluated as~\cite{Patti2012}:
\begin{equation*}
\Delta t_{B}=\frac{\mathcal{A}_i}{3} \Delta t_{MC},
\end{equation*}
where $ \mathcal{A}_i $ is  the  acceptance coefficient for the $i$-th MC step.

The total Brownian dynamics time was evaluated as the sum
\begin{equation}
 t_{B}=\frac{ \Delta t_{MC} } {3}  \sum_{i=1}^{t_{MC}}\mathcal{A}_i,
\end{equation}
where $ t_{MC} $ is  the  MC time.
\begin{figure}[!htbp]
	\centering
	\includegraphics[width=0.9\columnwidth]{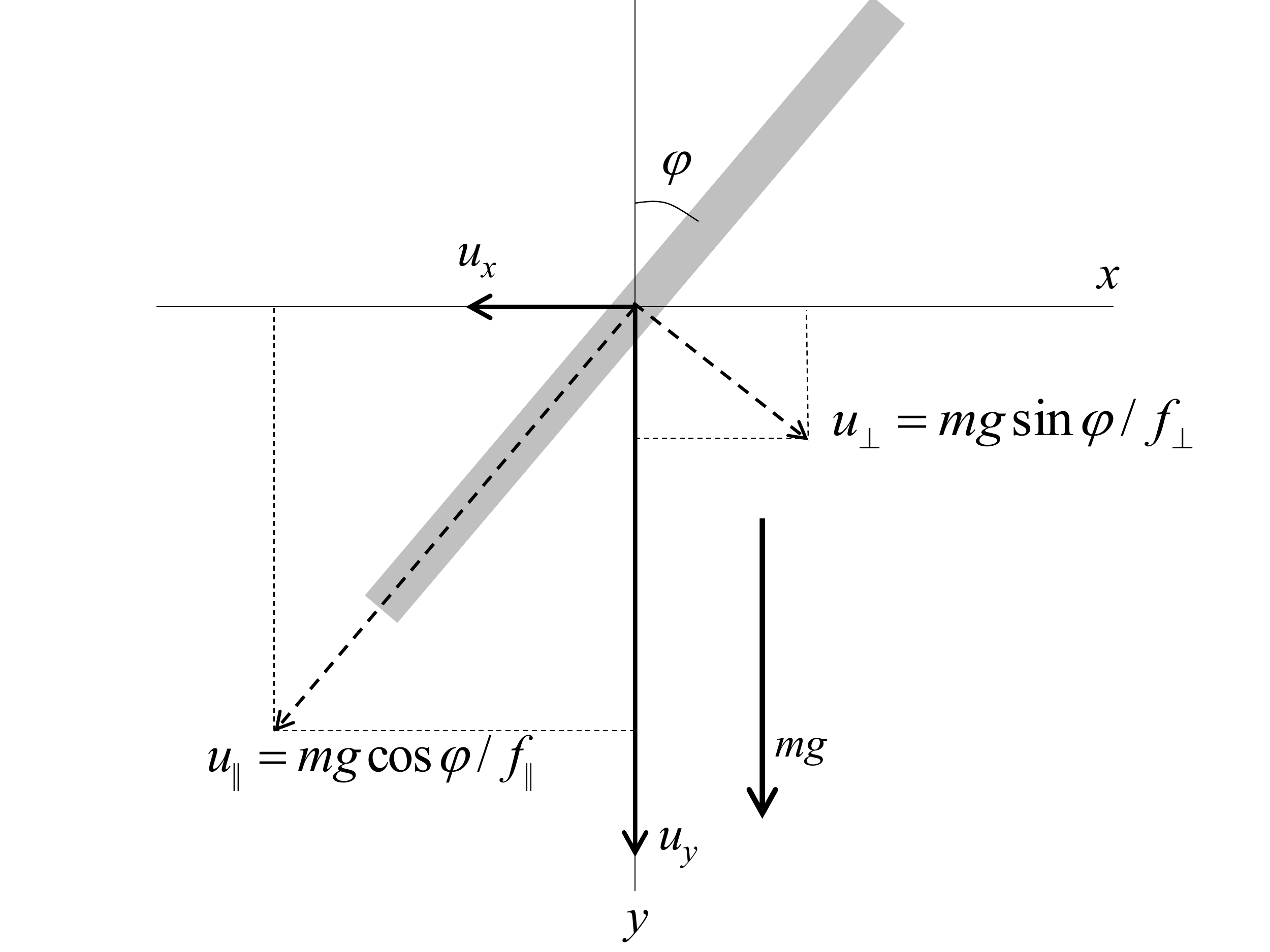}
	\caption{To the evaluation of the components of the sedimentation rates in the horizontal ($x$) and vertical ($y$) directions.   \label{fig:Formulas_uxy}}
\end{figure}

The sedimentation rate of a rod, $u$, depends upon the angle  $\varphi$ between the long axis and direction of the gravity (Fig.~\ref{fig:Formulas_uxy}). It can be estimated from a balance of the net sedimentation force with the opposing translational frictional force~\cite{Schuck2016}, and the maximum value is reached when $\varphi=0$
\begin{equation}\label{eq:Vel}
u=mg/f_\parallel.
\end{equation}
Here $m=\Delta\rho V$ is the buoyant mass of the rod, $V=\pi l^3/(4a^2)$ is the volume of the rod, $\Delta \rho$ is the difference between the densities of the particle and the liquid (compare with Eq.~\eqref{eq:u} for spherical particles).

In general case, when $\varphi \ne 0$, the sedimentation rates in the horizontal ($x$) and vertical ($y$) directions were different
\begin{subequations}\label{eq:uxy}
\begin{align}
u_x/u   &=  (-1+f_\parallel/f_\perp)\sin\varphi \cos \varphi, \\
u_y/u   &=  (1-f_\parallel/f_\perp)\cos^2\varphi +f_\parallel/f_\perp,
\end{align}
\end{subequations}
and at each step the displacements of the chosen particle were $\Delta x  =  l u_x / u $ and $\Delta y  =  l u_y / u$. Finally, the P\'{e}clet number can be evaluated as the ratio of the Brownian, $\tau_B$, and sedimentation, $\tau_S$, times required for the	motion of a rod over the distance of $l$, i.e.,
\begin{equation}\label{eq:Peu}
\mathrm{Pe}=\tau_B/\tau_S=u/\alpha^{2}.
\end{equation}

At the initial moment before settling the front of the settling layer coincides with the upper border and it moves in the downward direction in the course of the settling. Time counting was started from the value of $t_{MC}=1$, being the initial moment (before settling and diffusion), and the total duration of the simulation was typically $10^6$--$10^7$ MC time units.
\begin{figure}[!htbp]
  \centering
 \includegraphics[width=0.95\columnwidth]{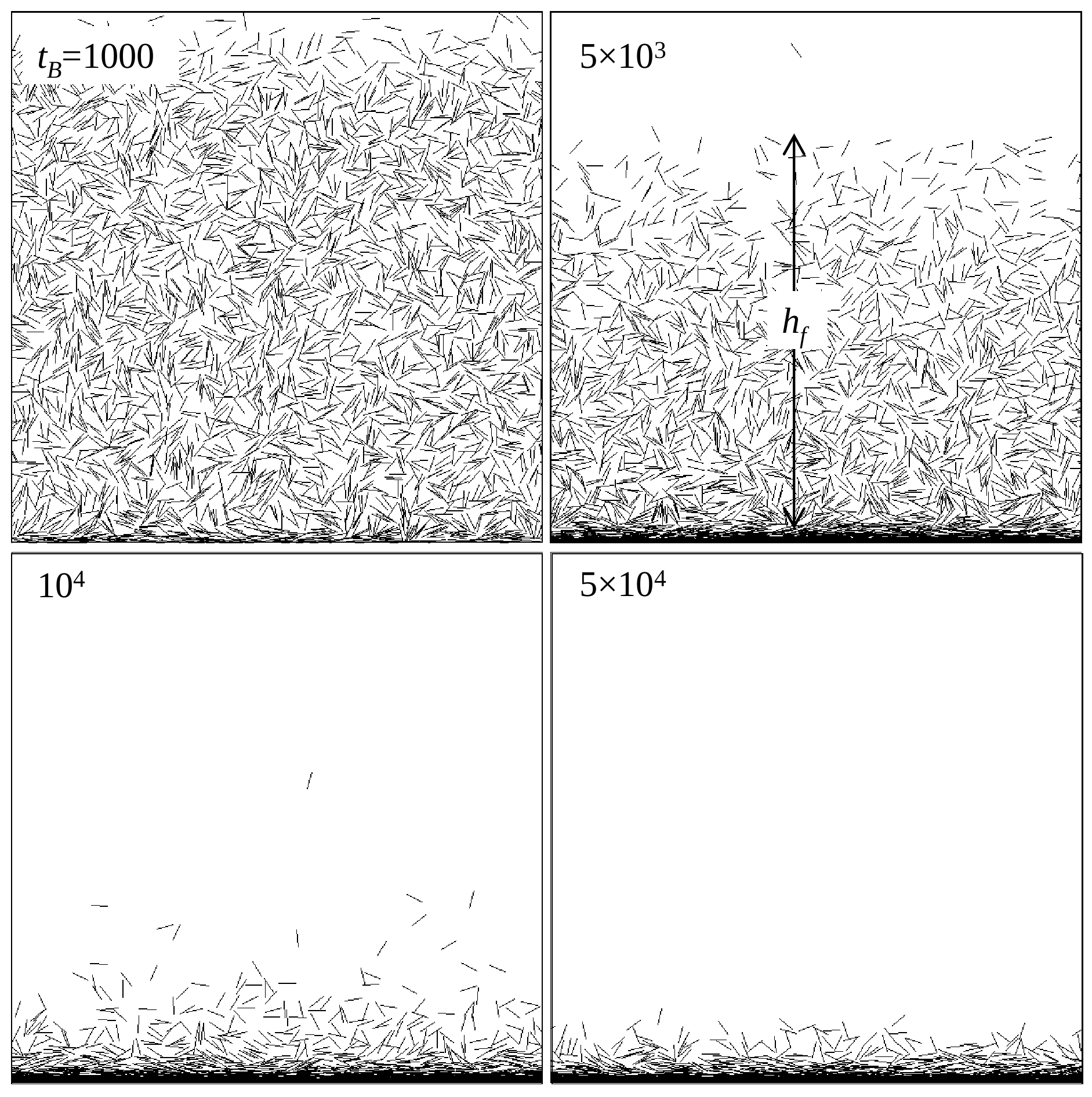}
  \caption{Example of the evolution of patterns at different sedimentation times, $t_B$, at sedimentation rate of $u=10^{-3}$ and initial number density of rods of $\rho_i=3$.\label{fig:Patternsf01}}
\end{figure}

Figure~\ref{fig:Patternsf01} presents an example of the evolution of the patterns at a sedimentation rate of $u=10^{-3}$ and an initial number density of $\rho_i=3$.  See Supplemental Material at [URL will be inserted by publisher] for an animation of the sedimentation for different values of the initial number density. To calculate the running height of the front of the settling layer, $H_f$, the film was divided along the vertical axis into regular columns with widths $L/32$, then the maximum value of the $y$ coordinates of all rods in each column was determined, and the value of $H_f$ was defined as the mean value averaged along all the columns. The relative height of the front of the settling layer was defined as $h_f=H_f/H$. The averaged number density of the particles in the settling layer with a height below $H_f$ can be calculated as $\rho_f=\rho_i / h_f$.

The mean order parameter in the settling layer was calculated as
\begin{equation}\label{eq:S}
S_f=\frac 1 N\sum\limits_{i=1}^{N} 2\cos^2 \theta_i -1,
\end{equation}
where $\theta_i$ is the angle between the axis of the $i$-th rod and the horizontal axis $x$, and $N$ is the total number of rods. To characterize the orientation of rods at different heights, we utilized $S(y)$, i.e., the order parameter~\eqref{eq:S} calculated within a narrow band $y, y + \Delta y$, where $\Delta y = l$. The profiles of the normalized number density, $\rho^*(y)=\rho(y)/\rho_i$, and the order parameter, $S(y)$, along the vertical axis $y$ were evaluated as the averaged values of $\rho/\rho_i$ and $S$ in the layers with height of $y$ ($y\leq H_f$).
\begin{figure}[!htbp]
  \centering
 \includegraphics[width=0.95\columnwidth]{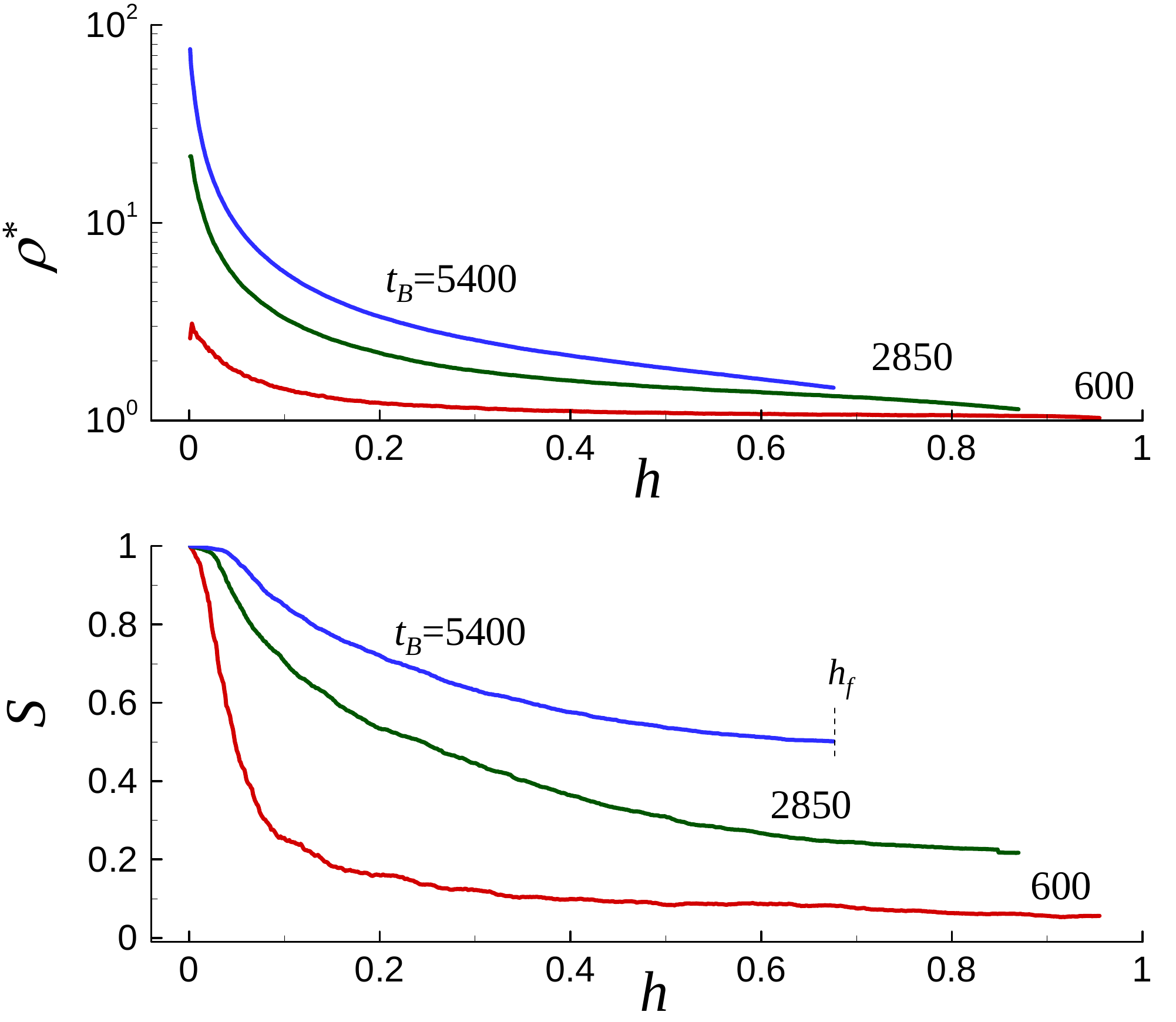}
  \caption{Example of the profiles of the normalized number density $\rho^*(h)=\rho(h)/\rho_i$, and order parameter, $S(h)$, along the vertical axis $y$ at different sedimentation times, $t_B$. Here, $h=y/H$ is a relative height. The sedimentation rate was $u=10^{-3}$ and the initial number density was $\rho_i=3$. \label{fig:Profilesf02}}
\end{figure}

Figure~\ref{fig:Profilesf02} presents an example of the profiles of the normalized number density, $\rho(h)/\rho_i$, and order parameter, $S(h)$, along the vertical axis $y$ at different sedimentation times, $t_B$. Here, $h = y / H$ is a relative height. These data were evaluated at a sedimentation rate of $u=10^{-3}$ and an initial number density of rods of $\rho_i=3$. At the initial moment of time ($t=1$) the distribution of rods in the layer was homogeneous with $\rho^*(h)=1$ and $S(h)=0$ inside the layer. In the course of sedimentation a densified layer is formed near the bottom with increased values of $\rho^*$ and $S$. The zone above $y>h_f$ corresponds to the supernatant without particles in~it.

To characterize the sediment film, the electrical conductivity at the end of simulation was calculated.
The  discretization scheme of the problem used a supporting mesh  with a mesh size  $L/m$ ($m \geq 256$)~\cite{Lebovka2018PREanisotropy}. The cells of the supporting mesh covered by rods were assumed to be conducting while the others were assumed to be insulating. In fact, the use of a supporting mesh  for the calculation of electrical conductivity is equivalent to rasterization of a structure of infinitely thin rods and the substitution of them by rods having a finite aspect ratio of the order of $a_m\approx m/L$. The system with infinitely thin rods corresponds to that with very large values of $m/L$.  However, the applied schema of discretization generates a ``zoo of lattice animals'', i.e., a set of polyominoes of different shapes and  sizes, especially, for small values of $m$. This set  is not complete because discretization of a rod cannot produce  polyominoes of all possible shapes~\cite{Lebovka2018PREanisotropy}. Therefore, we can only use this method for estimation of the behavior of electrical conductivity. Similar approaches have previously been applied for the estimation of electrical conductivities of aligned rods~\cite{Tarasevich2018PREb,Tarasevich2018JAP} and dried films of rods~\cite{Lebovka2018PREverticaldrying}.

In our work, the Frank---Lobb algorithm was applied to evaluate the electrical conductivity~\cite{Frank1988}. Note, that when the electrical conductivity of a host matrix is insignificant and the sticks are allowed to intersect, the utilization of Kirchhoff network approach can be also applied~\cite{Tarasevich2019}. We put $\sigma_i =1$, and $\sigma_c= 10^6$ in arbitrary units for the conductivities of the insulating and conducting sites, respectively. The bottom layers with different thicknesses, $\delta$, were selected, the two conducting buses were applied to the opposite borders, and the electrical conductivity was calculated between these buses in the horizontal, $\sigma_x$ and vertical $\sigma_y$ directions (see~\cite{Lebovka2016,Lebovka2017} for details).

For each given value of $\rho_i$ or $u$, the computer  experiments were repeated up to $100$ independent runs.  The error bars in the figures correspond to the standard deviations of the means. When not shown explicitly, they are of the order of the marker size.

\section{Results and Discussion\label{sec:results}}

\begin{figure}[!htbp]
  \centering
 \includegraphics[width=0.95\columnwidth]{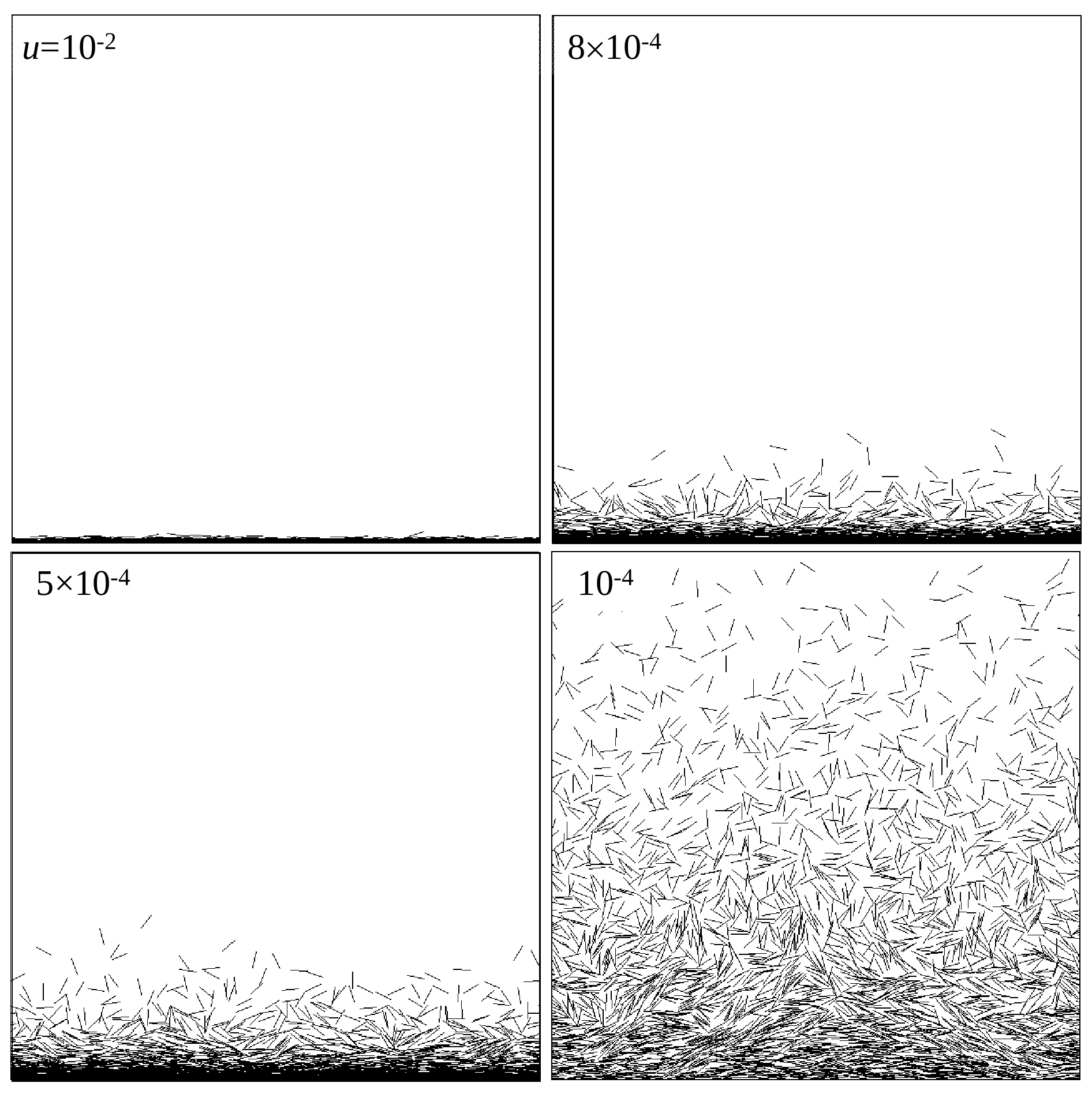}
  \caption{Sedimentation patterns at different sedimentation rates, $u$, for a fixed initial number density of $\rho_i=3$ and $t_B=2\times10^5$. \label{fig:PU}}
\end{figure}

Figure~\ref{fig:PU} compares sedimentation patterns at different sedimentation rates, $u$, for a fixed value of the initial number density of $\rho_i=3$ with relatively long time of sedimentation, $t_B=2\times10^5$. Similar patterns were also observed at other values of $\rho_i$ in the interval between 1 and 7.

Comparison of the patterns allows the following preliminary conclusion to be drawn. For relatively high sedimentation rates ($u= 10^{-2}$, $\mathrm{Pe}=4$), the formation of a compact bottom layer is observed. Its thickness increases with decreasing values of $u$ and at the smallest sedimentation rate  ($u= 10^{-4}$, $\mathrm{Pe}=4\times 10^{-2}$) no compact bottom layer can be observed owing to the presence of a sedimentation-diffusion equilibrium (SDE).

\begin{figure}[!htbp]
  \centering
 \includegraphics[width=0.95\columnwidth]{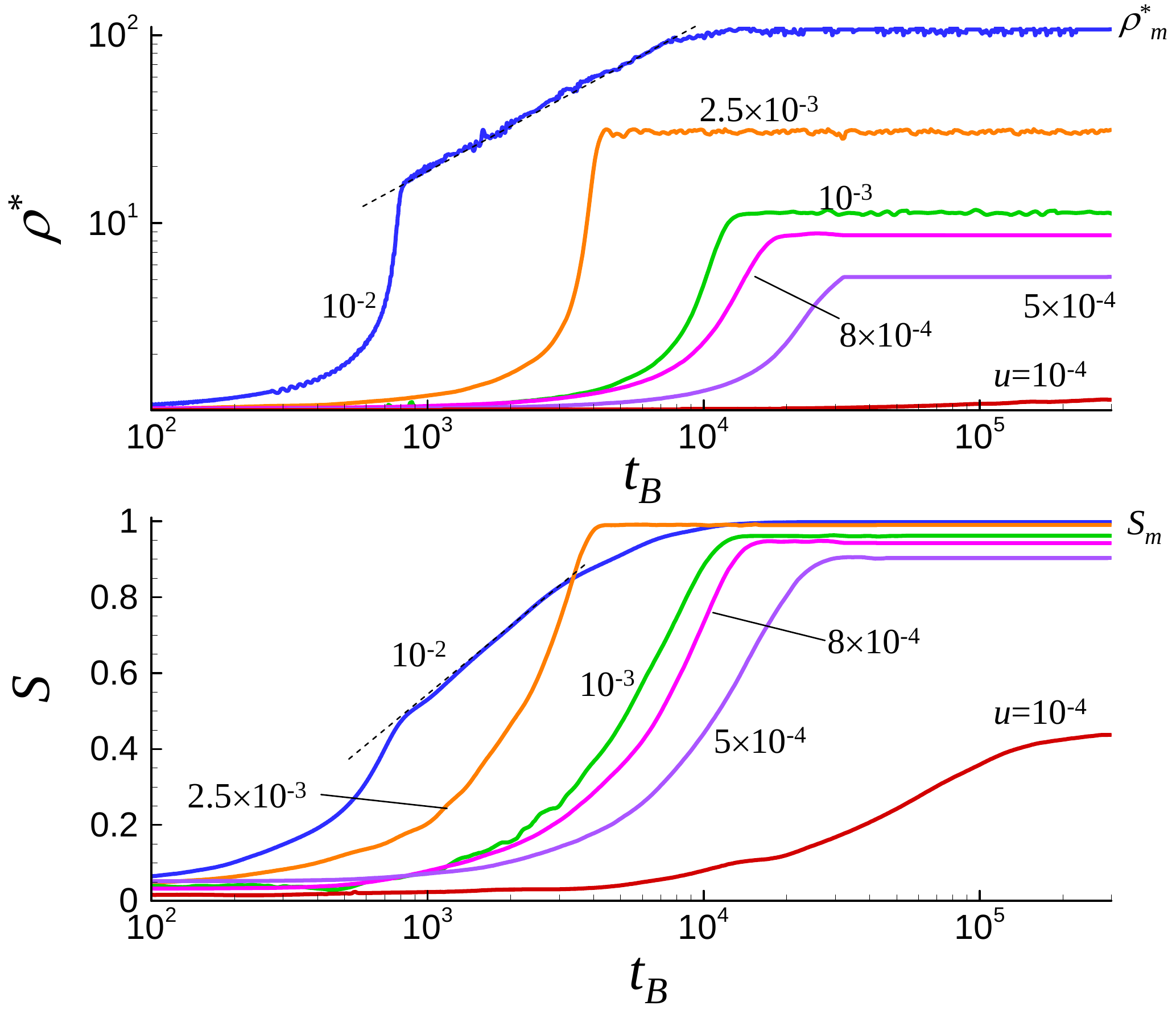}
  \caption{Evolution over time of the mean values of the normalized number density $\rho^*=\rho/\rho_i$ and the order parameter $S$ in sediment layers at different sedimentation rates, $u$, and an initial number density of rods of $\rho_i$. \label{fig:4Evol}}
\end{figure}

Figure~\ref{fig:4Evol} presents the mean normalized number density  $\rho^*$ and order parameter $S$ in the sediment layers versus the sedimentation time $t_B$ for the particular value of initial number density of rods of $\rho_i=3$.
For large sedimentation rates, $u= 10^{-2}$, $\mathrm{Pe}=4$, the sedimentation includes two steps with the formation of non-equilibrium porous film (a piled state with a porous structure and poorly oriented rods) at the initial stage ($t_B\approx 10^{2}-10^{3}$) followed by equilibration of the porous structure and the formation of compact highly oriented films ($S\approx 1$) subsequently ($t_B>10^{4}$).

At smaller sedimentation rates ($u<2.5\times10^{-3}$), only one stage of sedimentation was observed. The time of film formation $t_f$ increased with decrease of $u$. Moreover, the saturation level of the mean normalized number density $\rho^*_m=\rho_m/\rho_i$ and order parameter $S_m$ for longer sedimentation  times ($t_B\geq 5\times 10^{5}$) decreased with decreasing values of $u$. At some critical level below $u\approx 10^{-4}$ the value of $\rho^*_m$ tend to 1. That corresponded to the establishment of a SDE state (see Fig.~\ref{fig:PU}, $u\approx 10^{-4}$). Note that in the SDE state, the mean order parameter was non-zero, $S\approx 0.5$. It reflected the formation of dense oriented layers near the bottom of the sediment.

\begin{figure}[!htbp]
  \centering
 \includegraphics[width=0.95\columnwidth]{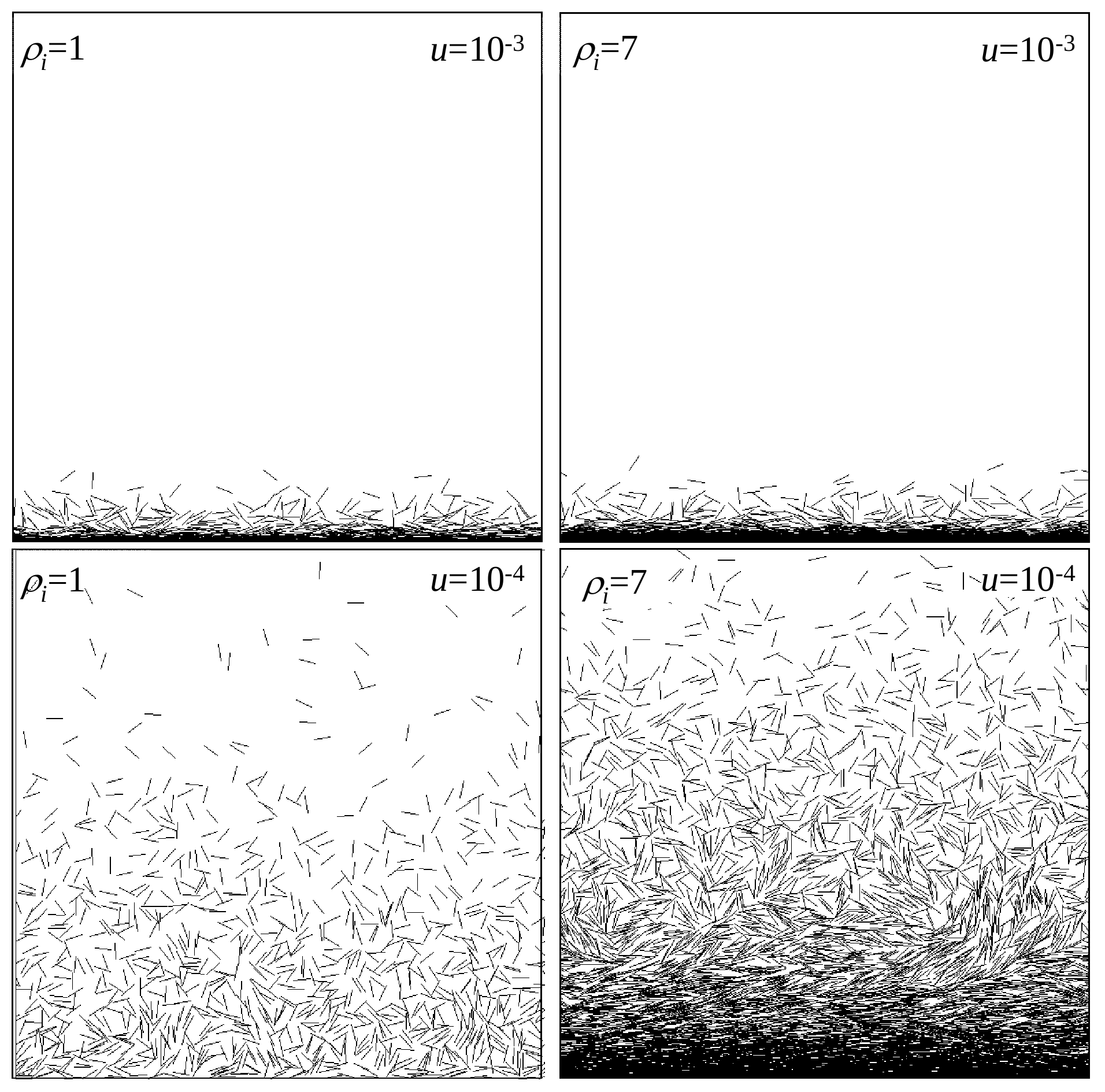}
  \caption{Example of sedimentation patterns at $t_B=10^6$ at two sedimentation rates ($u=10^{-3}$ and $u=10^{-4}$), and two initial number densities ($\rho_i=1$ and $\rho_i=7$). \label{fig:PatternsRho}}
\end{figure}

Similar dependencies, were observed for other studied values of $\rho_i$. The obtained data revealed some impact of the initial number density of rods $\rho_i$ on the sedimentation behavior. Figure~\ref{fig:PatternsRho} compares the sedimentation patterns for a long sedimentation time ($t_B=2\times10^5$, near saturation regime) for sedimentation rates ($u=10^{-3}$ and $u=10^{-4}$) and different initial number densities ($\rho_i=1$ and $\rho_i=7$). At the relatively high sedimentation rate of $u=10^{-3}$  ($\mathrm{Pe}=0.4$), compact bottom layers were formed for both the $\rho_i=1$ and $\rho_i=7$ concentrations. However, at the smaller sedimentation rate of $u=10^{-4}$  ($\mathrm{Pe}=0.04$), the SDE states for the $\rho_i=1$ and $\rho_i=9$ concentrations were rather different. For the smaller initial number density, $\rho_i=1$, the upper part of the system was fairly dilute, while at $\rho_i=9$ the rods occupied the whole volume and formed locally oriented structures near the bottom layer.

\begin{figure}[!htbp]
  \centering
 \includegraphics[width=0.95\columnwidth]{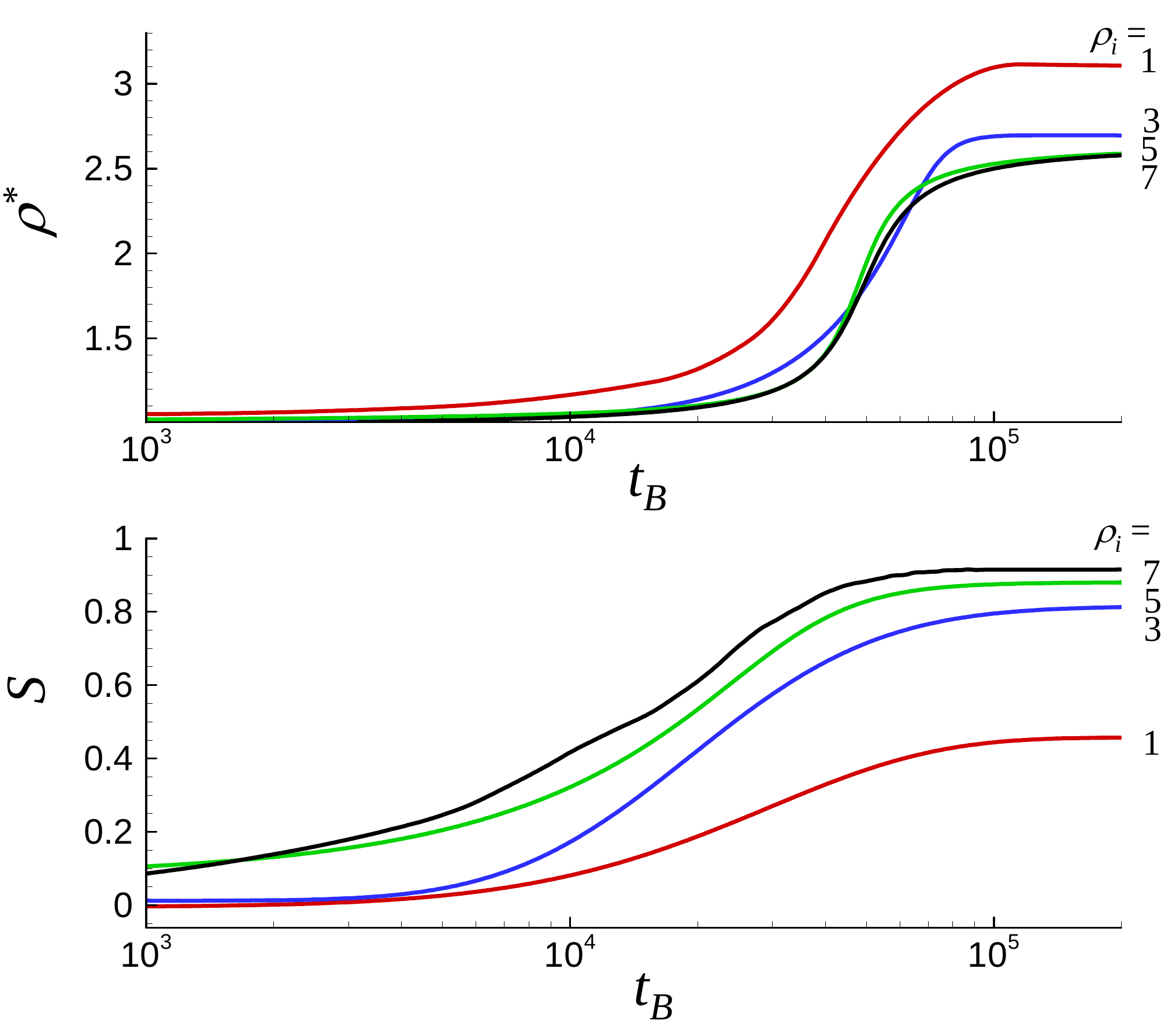}
  \caption{Examples of the evolution over time of the mean values of the normalized number density $\rho^*=\rho/\rho_i$ and the order parameter $S$ in sediment layers at a sedimentation rate of $u=5\times10^{-5}$, and different initial number density of rods $\rho_i$. \label{fig:EvolRo}}
\end{figure}

The initial number density $\rho_i^*$ also affects the kinetics of the mean values of the normalized number density $\rho^*=\rho/\rho_i$ and the order parameter $S$ in the sediment layers (Fig.~\ref{fig:EvolRo}). The presented data correspond to the near SDE regime ($\mathrm{Pe}=0.1$) in presence of the equilibrium between the sedimentation and diffusion processes. The data on $\rho^*(t_B)$ revealed that the saturated normalized number density $\rho^*$ decreased and the order parameter $S$ increased with increasing initial number density $\rho_i$.

\begin{figure}[!htbp]
  \centering
 \includegraphics[width=0.95\columnwidth]{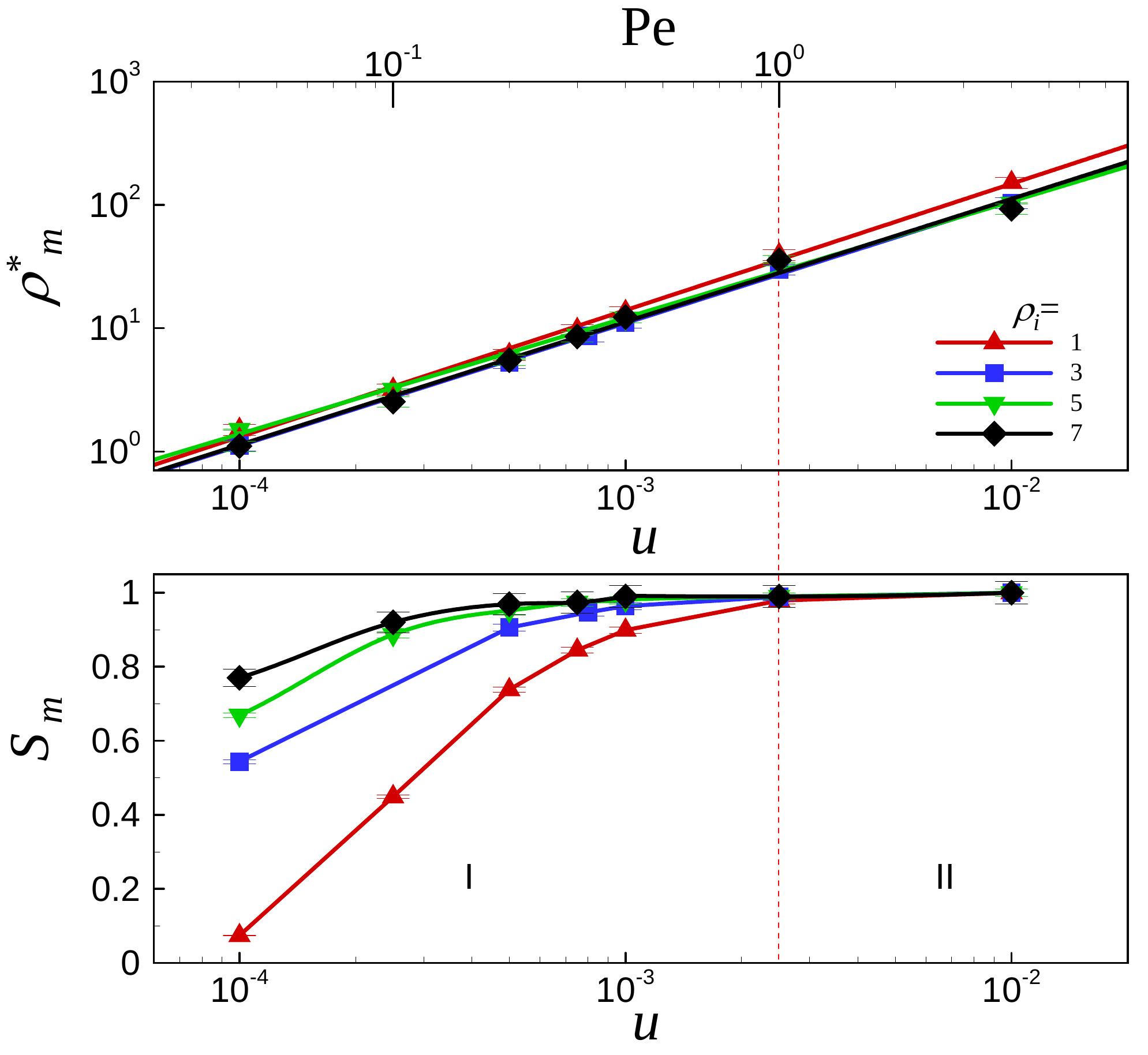}
  \caption{The maximum (saturated) values (at $t_B>3\times10^5$) of the normalized mean number density $\rho^*_m=\rho_m/\rho_i$ and the order parameter $S_m$ in sediment layers at different sedimentation rates, $u$, or values of $\mathrm{Pe}$  and initial number densities of rods, $\rho_i$. \label{fig:RomSm}}
\end{figure}

Figure~\ref{fig:RomSm} summarizes the observed dependencies of the saturated values of the normalized mean number density $\rho^*_m (u)$ and the order parameter $S_m (u)$ in sediment layers at different initial number densities, $\rho_i$. It is interesting that, across a wide range of sedimentation rates 
the $\rho^*_m(u)$ dependencies can be well fitted with high values of the coefficients of determination ($R^2>0.991$) by power relation
\begin{equation}\label{eq:Pow}
\rho^*_m\propto u^\beta,
\end{equation}
with the exponent of $\beta=0.987\pm 0.019$. Therefore, all these dependencies were almost linear over the investigated range of sedimentation rates.

However, structures with a high mean order parameter, $S\approx 1$ were only observed at large
values of $u$, $u\geq 2.5\times 10^{-3}$ ($\mathrm{Pe}\geq 1$) (Fig.~\ref{fig:RomSm}).
From these dependencies we can define the two zones of sedimentation rates that approximately correspond to the manifestation of the sedimentation-diffusion equilibrium state (zone~I, $\mathrm{Pe}< 1$) and formation of compact sediment films with high order parameter, $S\approx 1$, (zone~II, $\mathrm{Pe}> 1$). At very high sedimentation rates ($\mathrm{Pe}> 10$) formation of non-equilibrium piled states, i.e., porous films with poorly oriented rods were also observed (data are not presented).
\begin{figure}[!htbp]
  \centering
 \includegraphics[width=0.95\columnwidth]{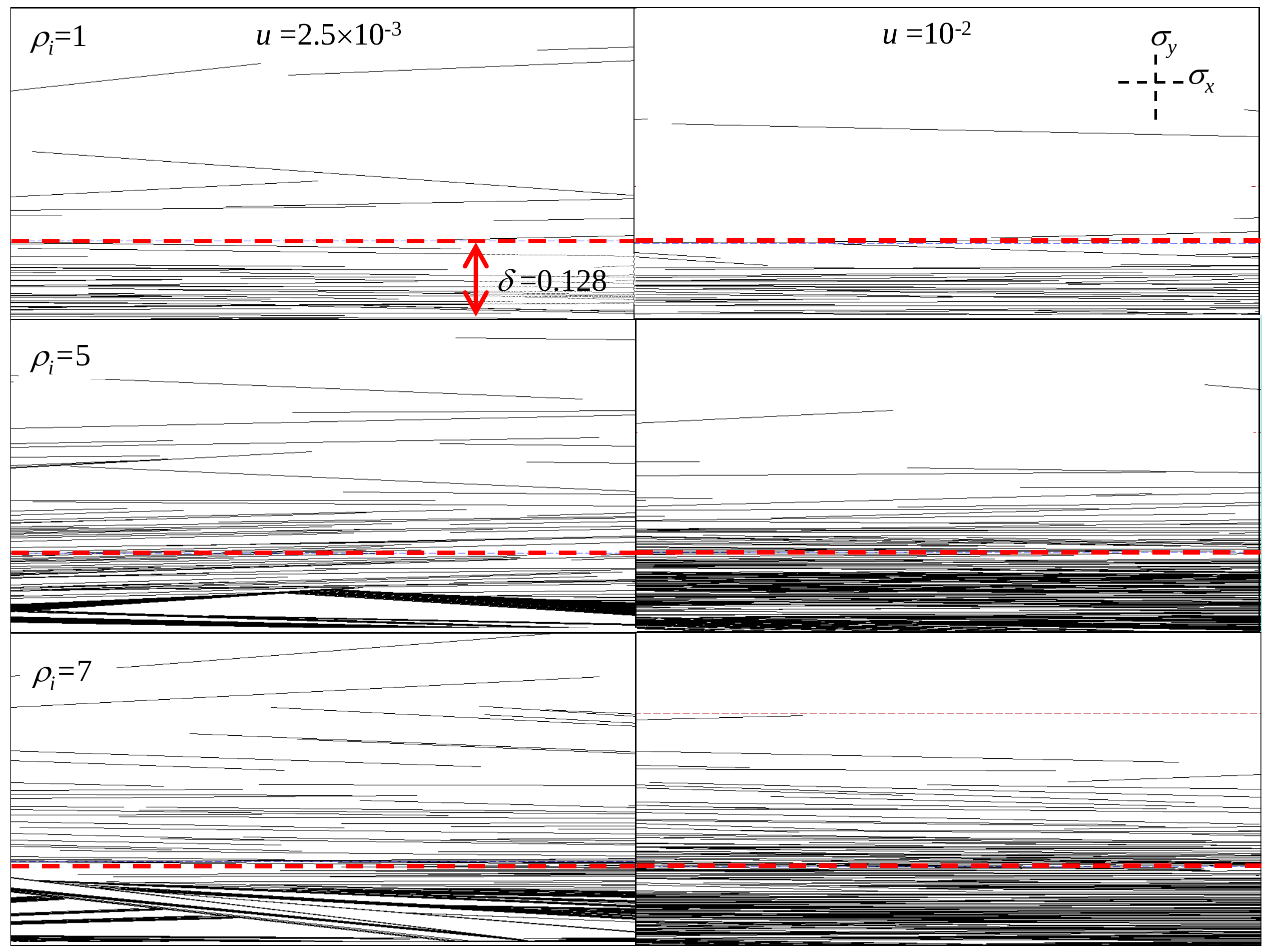}
  \caption{Enlarged portion of the sedimentation patterns with size of $l$ (horizontal) $\times 0.5l$ (vertical) at $t_B=3\times 10^5$ for different initial number density $\rho_i$, and two sedimentation rates ($u=2.5\times10^{-3}$ and $u=10^{-2}$). The horizontal dashed lines correspond to the bottom layers with the thickness of $\delta=0.128$. \label{fig:Patterns08}}
\end{figure}

The structure of the sediment at the boundary between zones I and II can significantly depend upon the initial number density of the rods $\rho_i$. Figure~\ref{fig:Patterns08}  presents the enlarged portions of the sedimentation patterns at $t_B = 3\times 10^5$ for different initial number densities $\rho_i$ for two 
sedimentation rates $u = 2.5 \times 10^{-3}$ (at the boundary between zones I and II, $\mathrm{Pe}=1$ and $u =10^{-2}$ ($\mathrm{Pe}=4$). An increase in $\rho_i$ resulted in an increase of the height of the sediments at both sedimentation rates. However, for large initial densities $\rho_i>5$,  significant stack-like porous structures could be clearly observed for $u =2.5\times 10^{-3}$ but these were absent for $u = 10^{-2}$.
\begin{figure}[!htbp]
\centering
  \includegraphics[width=0.9\columnwidth]{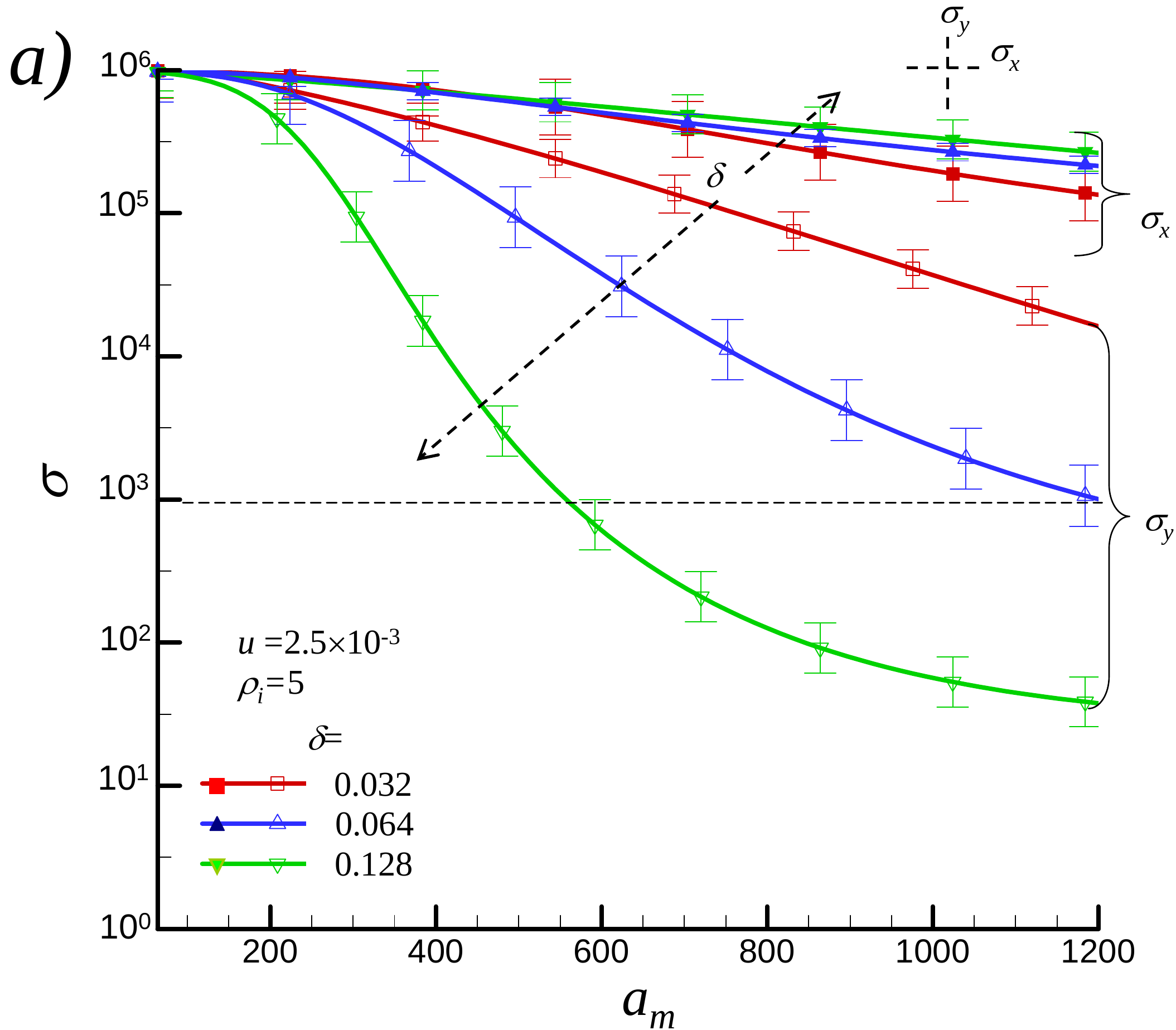}\\
  \includegraphics[width=0.9\columnwidth]{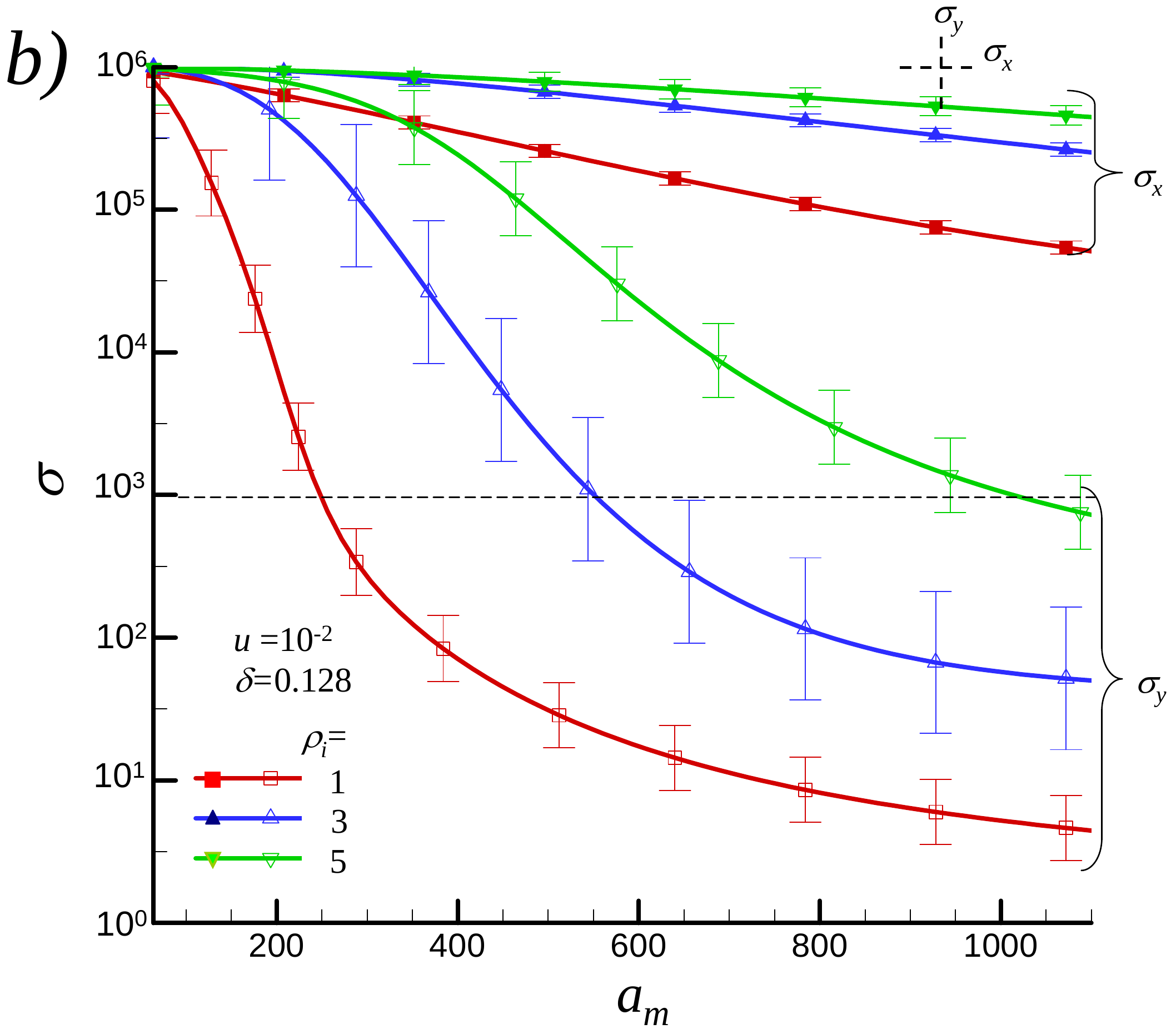}\\
\caption{Electrical conductivity (arbitrary units) in the horizontal, $\sigma_x$, and vertical, $\sigma_y$, directions versus the mesh aspect ratio of rods, $a_m$, at different thicknesses of the bottom layers, $\delta$, ($\rho_i=5$, $u=2.5\times10^{-3}$ (a) and different initial number densities $\rho_i$, ($u=10^{-2}$, $\delta=0.128$) (b), and $t_B=3\times 10^5$. \label{fig:Conduct}}
\end{figure}

Such differences in sediment film structures can affect their physical properties. We performed tests of electrical conductivity on compact sediment films with high order parameters at sedimentation rates ($u=2.5\times10^{-3}$ and $u=10^{-2}$).
Figure~\ref{fig:Conduct} shows the electrical conductivity in the horizontal $x$ and vertical $y$ directions versus the mesh aspect ratio of rods, $a_m$, defined by the size of the supporting mesh. The data are presented at $t_B = 3\times 10^5$ with examples for fixed initial number density $\rho=5$ and sedimentation rate $u=2.5\times10^{-3}$, and different thicknesses of the bottom  layers $\delta$ (Fig.~\ref{fig:Conduct}a) and fixed $\delta$ ($\delta=0.128$) and $u=10^{-2}$, and different initial number densities over the interval $\rho \in [1,7]$ (Fig.~\ref{fig:Conduct}b).

The similar tendencies in the $\sigma (a_m)$ dependencies were observed in both the $x$ and $y$ directions. At relatively small values of $a_m$ $(< 100)$, the electrical conductivity was fairly high above the level of the mean geometrical conductivity defined as $\sigma_g=\sqrt{\sigma_i\sigma_c}=10^3$. We can treat a system with conductivity $\sigma > \sigma_g$ as conducting while a system with conductivity $\sigma < \sigma_g$ can be considered to be  insulating~\cite{Lebovka2018PREanisotropy,Tarasevich2018PREb}. The values of $\sigma_x$ and $\sigma_y$ continuously decreased with increasing $a_m$, while loss of percolation at electrical conductivities smaller than the value of $\sigma_g$ can occur at large values of $a_m$. This simply reflect the damage of the connectivity inside the mesh clusters with thinning of the mesh structure (increasing $a_m$).

At fixed $a_m$ the increase of $\delta$ resulted in decrease of $\sigma_y$ and increase of $\sigma_x$, i.e., anisotropy of electrical conductivity became larger (Fig.~\ref{fig:Conduct}a). It can reflect the changes in connectivity for different thicknesses of the bottom layers.
\begin{figure}[!htbp]
  \centering
 \includegraphics[width=0.95\columnwidth]{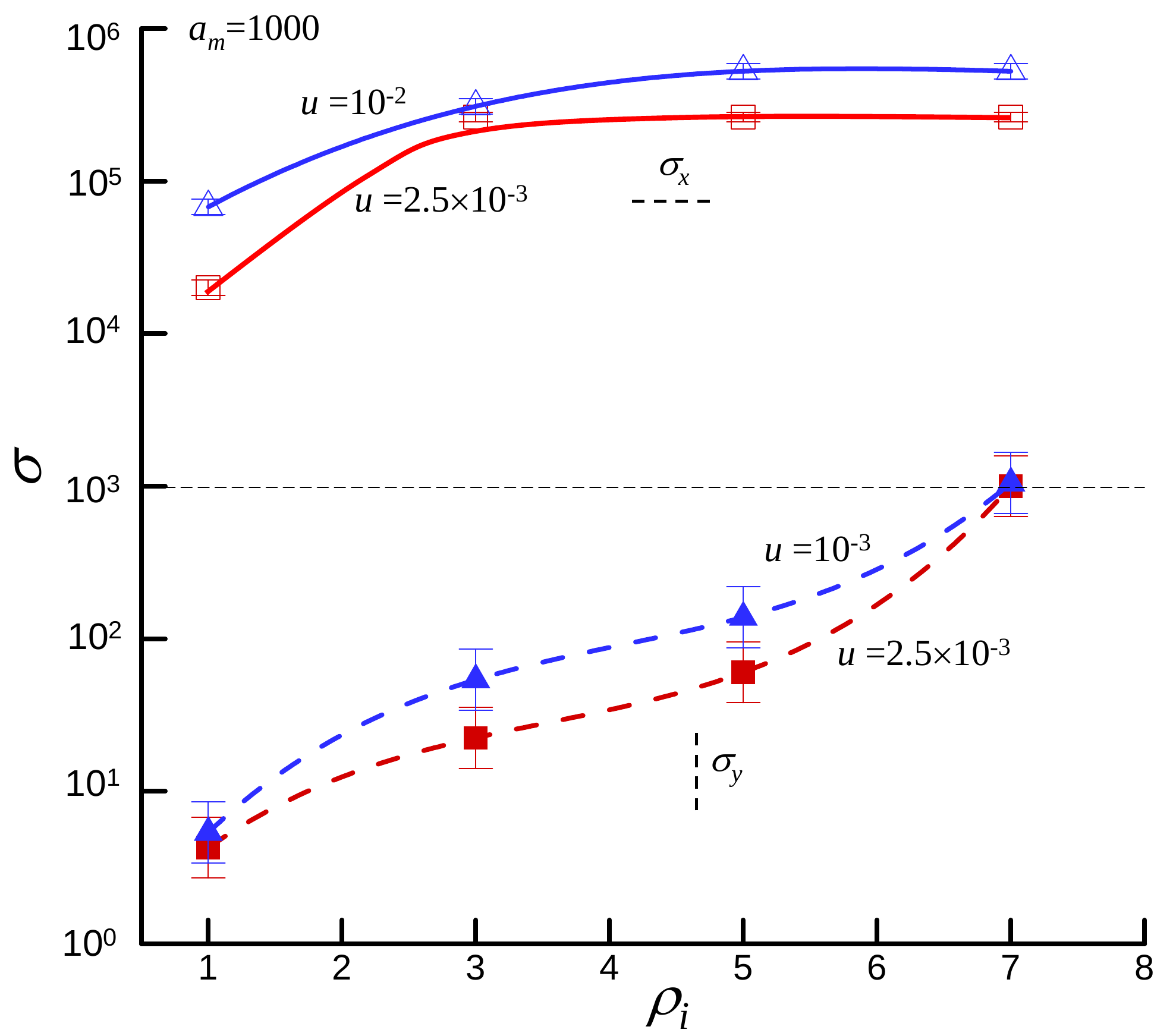}
  \caption{Electrical conductivity (arbitrary units) in the horizontal, $\sigma_x$, and vertical, $\sigma_y$, directions versus the initial number density of rods $\rho_i$ at two sedimentation rates ($u=5\times10^{-4}$ and $u=10^{-3}$), $t=10^6$ and $a_m=900$.
	\label{fig:ElCond}}
\end{figure}

Figure~\ref{fig:ElCond} shows examples of the $\sigma (\rho_i)$ dependencies in the horizontal and vertical directions at fixed $a_m = 1000$, $\delta=0.128$ and two sedimentation rates ($u=2.5\times10^{-3}$ and $u=10^{-2}$). In the horizontal direction $x$, the electrical conductivity $\sigma_x$ exceeded the values of $\sigma_g$ even at small initial densities $\rho_i> 1$. In the vertical direction $y$, the values of $\sigma_y$ were approaching $\sigma_g$ at $\rho_i=7$. Moreover, the increase in the sedimentation rate resulted in noticeable increase of both the $\sigma_x$ and $\sigma_y$. 	This evidently reflected the differences in the sediment film structure displayed in Fig.~\ref{fig:Patterns08}.

\section{Conclusion\label{sec:conclusion}}
A continuous 2D model of the sedimentation in suspension of rods was studied using MC simulation. During the sedimentation, the rods underwent translational and rotational Brownian motions. Significant sedimentation-driven self-assembly of the rods was observed. Different steps were observed, with the formation of non-equilibrium porous films during the initial stage followed by equilibration and the formation of compact highly oriented films over  longer sedimentation time. Two zones of sedimentation rate were observed approximately corresponding to the manifestation of the sedimentation-diffusion equilibrium state  (zone~I, $\mathrm{Pe}< 1$); the formation of compact sediment films with high order parameter, $S \approx 1$, (zone~II, $\mathrm{Pe}> 1$). Within zones I and II the densities of the sediment films followed the power relation $\rho^*_m\propto u^\beta$ with the exponent of $\beta=0.987\pm 0.019$.

This study provides the scientific community with information regarding the structure of sediment films of conducting rods, as well as how the properties of these films depend on the sedimentation conditions. The simulation of the electrical conductivity in the horizontal $x$ and vertical $y$ directions evidenced that the electrical conductivity of the resulting sediment films displays high anisotropy with $\sigma_x > \sigma_y$. The degree of anisotropy can be regulated by changing the rate of sedimentation, $u$, and the initial number density of the rods in suspension,~$\rho_i$.

\section*{Acknowledgements}
We acknowledge the funding from the National Academy of Sciences of Ukraine, Projects No.~0117U004046 and 43/19-H (N.I.L., N.V.V.), and the Ministry of Science and Higher Education of the Russian Federation, Project No.~3.959.2017/4.6 (Yu.Yu.T.). The simulations for this work were partially performed using the computing cluster of the Taras Shevchenko National University of Kiev\footnote{\url{http://cluster.univ.kiev.ua/eng/}}.

\bibliography{Settling}

\begin{thebibliography}{68}%
\makeatletter
\providecommand \@ifxundefined [1]{%
 \@ifx{#1\undefined}
}%
\providecommand \@ifnum [1]{%
 \ifnum #1\expandafter \@firstoftwo
 \else \expandafter \@secondoftwo
 \fi
}%
\providecommand \@ifx [1]{%
 \ifx #1\expandafter \@firstoftwo
 \else \expandafter \@secondoftwo
 \fi
}%
\providecommand \natexlab [1]{#1}%
\providecommand \enquote  [1]{``#1''}%
\providecommand \bibnamefont  [1]{#1}%
\providecommand \bibfnamefont [1]{#1}%
\providecommand \citenamefont [1]{#1}%
\providecommand \href@noop [0]{\@secondoftwo}%
\providecommand \href [0]{\begingroup \@sanitize@url \@href}%
\providecommand \@href[1]{\@@startlink{#1}\@@href}%
\providecommand \@@href[1]{\endgroup#1\@@endlink}%
\providecommand \@sanitize@url [0]{\catcode `\\12\catcode `\$12\catcode
  `\&12\catcode `\#12\catcode `\^12\catcode `\_12\catcode `\%12\relax}%
\providecommand \@@startlink[1]{}%
\providecommand \@@endlink[0]{}%
\providecommand \url  [0]{\begingroup\@sanitize@url \@url }%
\providecommand \@url [1]{\endgroup\@href {#1}{\urlprefix }}%
\providecommand \urlprefix  [0]{URL }%
\providecommand \Eprint [0]{\href }%
\providecommand \doibase [0]{http://dx.doi.org/}%
\providecommand \selectlanguage [0]{\@gobble}%
\providecommand \bibinfo  [0]{\@secondoftwo}%
\providecommand \bibfield  [0]{\@secondoftwo}%
\providecommand \translation [1]{[#1]}%
\providecommand \BibitemOpen [0]{}%
\providecommand \bibitemStop [0]{}%
\providecommand \bibitemNoStop [0]{.\EOS\space}%
\providecommand \EOS [0]{\spacefactor3000\relax}%
\providecommand \BibitemShut  [1]{\csname bibitem#1\endcsname}%
\let\auto@bib@innerbib\@empty
\bibitem [{\citenamefont {Schaflinger}(1990)}]{Schaflinger1990}%
  \BibitemOpen
  \bibfield  {author} {\bibinfo {author} {\bibfnamefont {Uwe}\ \bibnamefont
  {Schaflinger}},\ }\bibfield  {title} {\enquote {\bibinfo {title} {Centrifugal
  separation of a mixture},}\ }\href {\doibase 10.1016/0169-5983(90)90014-P}
  {\bibfield  {journal} {\bibinfo  {journal} {Fluid Dyn. Res.}\ }\textbf
  {\bibinfo {volume} {6}},\ \bibinfo {pages} {213} (\bibinfo {year}
  {1990})}\BibitemShut {NoStop}%
\bibitem [{\citenamefont {Tiller}\ \emph {et~al.}(1995)\citenamefont {Tiller},
  \citenamefont {Hsyung},\ and\ \citenamefont {Cong}}]{Tiller1995}%
  \BibitemOpen
  \bibfield  {author} {\bibinfo {author} {\bibfnamefont {Frank~M.}\
  \bibnamefont {Tiller}}, \bibinfo {author} {\bibfnamefont {N.~B.}\
  \bibnamefont {Hsyung}}, \ and\ \bibinfo {author} {\bibfnamefont {D.~Z.}\
  \bibnamefont {Cong}},\ }\bibfield  {title} {\enquote {\bibinfo {title} {Role
  of porosity in filtration: {XII.} {Filtration} with sedimentation},}\ }\href
  {\doibase 10.1002/aic.690410511} {\bibfield  {journal} {\bibinfo  {journal}
  {{AIChE} J.}\ }\textbf {\bibinfo {volume} {41}},\ \bibinfo {pages}
  {1153--1164} (\bibinfo {year} {1995})}\BibitemShut {NoStop}%
\bibitem [{\citenamefont {Chen}\ and\ \citenamefont {Scott}(1997)}]{Chen1997}%
  \BibitemOpen
  \bibfield  {author} {\bibinfo {author} {\bibfnamefont {Wu}~\bibnamefont
  {Chen}}\ and\ \bibinfo {author} {\bibfnamefont {Keith~J.}\ \bibnamefont
  {Scott}},\ }\bibfield  {title} {\enquote {\bibinfo {title} {Sedimentation},}\
  }in\ \href {\doibase 10.1007/978-1-4615-6373-0_13} {\emph {\bibinfo
  {booktitle} {Handbook of Powder Science \& Technology}}}\ (\bibinfo
  {publisher} {Springer},\ \bibinfo {year} {1997})\ pp.\ \bibinfo {pages}
  {635--682}\BibitemShut {NoStop}%
\bibitem [{\citenamefont {Chen}\ \emph {et~al.}(2015)\citenamefont {Chen},
  \citenamefont {C\"{o}lfen},\ and\ \citenamefont {Polarz}}]{Chen2015}%
  \BibitemOpen
  \bibfield  {author} {\bibinfo {author} {\bibfnamefont {Mengdi}\ \bibnamefont
  {Chen}}, \bibinfo {author} {\bibfnamefont {Helmut}\ \bibnamefont
  {C\"{o}lfen}}, \ and\ \bibinfo {author} {\bibfnamefont {Sebastian}\
  \bibnamefont {Polarz}},\ }\bibfield  {title} {\enquote {\bibinfo {title}
  {Centrifugal field-induced colloidal assembly: From chaos to order},}\ }\href
  {\doibase 10.1021/acsnano.5b01116} {\bibfield  {journal} {\bibinfo  {journal}
  {ACS Nano}\ }\textbf {\bibinfo {volume} {9}},\ \bibinfo {pages} {6944--6950}
  (\bibinfo {year} {2015})}\BibitemShut {NoStop}%
\bibitem [{\citenamefont {Petek}\ \emph {et~al.}(2015)\citenamefont {Petek},
  \citenamefont {Bukovec},\ and\ \citenamefont {{\v{S}}kofic}}]{Petek2015}%
  \BibitemOpen
  \bibfield  {author} {\bibinfo {author} {\bibfnamefont {Ur{\v{s}}a}\
  \bibnamefont {Petek}}, \bibinfo {author} {\bibfnamefont {Peter}\ \bibnamefont
  {Bukovec}}, \ and\ \bibinfo {author} {\bibfnamefont {Irena~Kozjek}\
  \bibnamefont {{\v{S}}kofic}},\ }\bibfield  {title} {\enquote {\bibinfo
  {title} {Preparation of electrically conductive au thin films by colloid
  sedimentation},}\ }\href {\doibase 10.17344/acsi.2014.1056} {\bibfield
  {journal} {\bibinfo  {journal} {Acta Chimica Slovenica}\ }\textbf {\bibinfo
  {volume} {62}},\ \bibinfo {pages} {281--287} (\bibinfo {year}
  {2015})}\BibitemShut {NoStop}%
\bibitem [{\citenamefont {Perrin}(1913)}]{Perrin1913}%
  \BibitemOpen
  \bibfield  {author} {\bibinfo {author} {\bibfnamefont {Jean~Baptiste}\
  \bibnamefont {Perrin}},\ }\href@noop {} {\emph {\bibinfo {title} {Les
  Atomes}}}\ (\bibinfo  {publisher} {Felix Alcan, Paris},\ \bibinfo {year}
  {1913})\BibitemShut {NoStop}%
\bibitem [{\citenamefont {Schuck}(2016)}]{Schuck2016}%
  \BibitemOpen
  \bibfield  {author} {\bibinfo {author} {\bibfnamefont {Peter}\ \bibnamefont
  {Schuck}},\ }\href {\doibase 10.1201/9781315367231} {\emph {\bibinfo {title}
  {Sedimentation Velocity Analytical Ultracentrifugation: Discrete Species and
  Size-Distributions of Macromolecules and Particles}}}\ (\bibinfo  {publisher}
  {CRC Press},\ \bibinfo {year} {2016})\BibitemShut {NoStop}%
\bibitem [{\citenamefont {Royall}\ \emph {et~al.}(2007)\citenamefont {Royall},
  \citenamefont {Dzubiella}, \citenamefont {Schmidt},\ and\ \citenamefont {van
  Blaaderen}}]{Royall2007}%
  \BibitemOpen
  \bibfield  {author} {\bibinfo {author} {\bibfnamefont {C.~Patrick}\
  \bibnamefont {Royall}}, \bibinfo {author} {\bibfnamefont {Joachim}\
  \bibnamefont {Dzubiella}}, \bibinfo {author} {\bibfnamefont {Matthias}\
  \bibnamefont {Schmidt}}, \ and\ \bibinfo {author} {\bibfnamefont {Alfons}\
  \bibnamefont {van Blaaderen}},\ }\bibfield  {title} {\enquote {\bibinfo
  {title} {Nonequilibrium sedimentation of colloids on the particle scale},}\
  }\href {\doibase 10.1103/PhysRevLett.98.188304} {\bibfield  {journal}
  {\bibinfo  {journal} {Phys. Rev. Lett.}\ }\textbf {\bibinfo {volume} {98}},\
  \bibinfo {pages} {188304} (\bibinfo {year} {2007})}\BibitemShut {NoStop}%
\bibitem [{\citenamefont {Thies-Weesie}\ \emph {et~al.}(1995)\citenamefont
  {Thies-Weesie}, \citenamefont {Philipse}, \citenamefont {N{\"a}gele},
  \citenamefont {Mandl},\ and\ \citenamefont {Klein}}]{Thies-Weesie1995}%
  \BibitemOpen
  \bibfield  {author} {\bibinfo {author} {\bibfnamefont {Dominique M.~E.}\
  \bibnamefont {Thies-Weesie}}, \bibinfo {author} {\bibfnamefont {Albert~P.}\
  \bibnamefont {Philipse}}, \bibinfo {author} {\bibfnamefont {Gerhard}\
  \bibnamefont {N{\"a}gele}}, \bibinfo {author} {\bibfnamefont {Barbara}\
  \bibnamefont {Mandl}}, \ and\ \bibinfo {author} {\bibfnamefont {Rudolf}\
  \bibnamefont {Klein}},\ }\bibfield  {title} {\enquote {\bibinfo {title}
  {Nonanalytical concentration dependence of sedimentation of charged silica
  spheres in an organic solvent: experiments and calculations},}\ }\href
  {\doibase 10.1006/jcis.1995.0006} {\bibfield  {journal} {\bibinfo  {journal}
  {J. Colloid Interface Sci.}\ }\textbf {\bibinfo {volume} {176}},\ \bibinfo
  {pages} {43--54} (\bibinfo {year} {1995})}\BibitemShut {NoStop}%
\bibitem [{\citenamefont {Segre}\ \emph {et~al.}(2001)\citenamefont {Segre},
  \citenamefont {Liu}, \citenamefont {Umbanhowar},\ and\ \citenamefont
  {Weitz}}]{Segre2001}%
  \BibitemOpen
  \bibfield  {author} {\bibinfo {author} {\bibfnamefont {Philip~N.}\
  \bibnamefont {Segre}}, \bibinfo {author} {\bibfnamefont {Fang}\ \bibnamefont
  {Liu}}, \bibinfo {author} {\bibfnamefont {P.}~\bibnamefont {Umbanhowar}}, \
  and\ \bibinfo {author} {\bibfnamefont {David~A.}\ \bibnamefont {Weitz}},\
  }\bibfield  {title} {\enquote {\bibinfo {title} {An effective gravitational
  temperature for sedimentation},}\ }\href {\doibase 10.1038/35054518}
  {\bibfield  {journal} {\bibinfo  {journal} {Nature}\ }\textbf {\bibinfo
  {volume} {409}},\ \bibinfo {pages} {594--597} (\bibinfo {year}
  {2001})}\BibitemShut {NoStop}%
\bibitem [{\citenamefont {Padding}\ and\ \citenamefont
  {Louis}(2004)}]{Padding2004}%
  \BibitemOpen
  \bibfield  {author} {\bibinfo {author} {\bibfnamefont {J.~T.}\ \bibnamefont
  {Padding}}\ and\ \bibinfo {author} {\bibfnamefont {A.~A.}\ \bibnamefont
  {Louis}},\ }\bibfield  {title} {\enquote {\bibinfo {title} {Hydrodynamic and
  {Brownian} fluctuations in sedimenting suspensions},}\ }\href {\doibase
  10.1103/PhysRevLett.93.220601} {\bibfield  {journal} {\bibinfo  {journal}
  {Phys. Rev. Lett.}\ }\textbf {\bibinfo {volume} {93}},\ \bibinfo {pages}
  {220601} (\bibinfo {year} {2004})}\BibitemShut {NoStop}%
\bibitem [{\citenamefont {Wachs}(2009)}]{Wachs2009}%
  \BibitemOpen
  \bibfield  {author} {\bibinfo {author} {\bibfnamefont {Anthony}\ \bibnamefont
  {Wachs}},\ }\bibfield  {title} {\enquote {\bibinfo {title} {A {DEM-DLM/FD}
  method for direct numerical simulation of particulate flows: Sedimentation of
  polygonal isometric particles in a {Newtonian} fluid with collisions},}\
  }\href {\doibase 10.1016/j.compfluid.2009.01.005} {\bibfield  {journal}
  {\bibinfo  {journal} {Computers \& Fluids}\ }\textbf {\bibinfo {volume}
  {38}},\ \bibinfo {pages} {1608--1628} (\bibinfo {year} {2009})}\BibitemShut
  {NoStop}%
\bibitem [{\citenamefont {Choi}\ \emph {et~al.}(2013)\citenamefont {Choi},
  \citenamefont {Yoon},\ and\ \citenamefont {Ha}}]{Choi2013}%
  \BibitemOpen
  \bibfield  {author} {\bibinfo {author} {\bibfnamefont {Changyoung}\
  \bibnamefont {Choi}}, \bibinfo {author} {\bibfnamefont {Hyun~Sik}\
  \bibnamefont {Yoon}}, \ and\ \bibinfo {author} {\bibfnamefont {Man~Yeong}\
  \bibnamefont {Ha}},\ }\bibfield  {title} {\enquote {\bibinfo {title} {Flow
  and motion characteristics of a freely falling square particle in a
  channel},}\ }\href {\doibase 10.1016/j.compfluid.2013.02.019} {\bibfield
  {journal} {\bibinfo  {journal} {Computers \& Fluids}\ }\textbf {\bibinfo
  {volume} {79}},\ \bibinfo {pages} {1--12} (\bibinfo {year}
  {2013})}\BibitemShut {NoStop}%
\bibitem [{\citenamefont {Karimnejad}\ \emph {et~al.}(2018)\citenamefont
  {Karimnejad}, \citenamefont {Delouei}, \citenamefont {Nazari}, \citenamefont
  {Shahmardan},\ and\ \citenamefont {Mohamad}}]{Karimnejad2018}%
  \BibitemOpen
  \bibfield  {author} {\bibinfo {author} {\bibfnamefont {S.}~\bibnamefont
  {Karimnejad}}, \bibinfo {author} {\bibfnamefont {A.~Amiri}\ \bibnamefont
  {Delouei}}, \bibinfo {author} {\bibfnamefont {M.}~\bibnamefont {Nazari}},
  \bibinfo {author} {\bibfnamefont {M.~M.}\ \bibnamefont {Shahmardan}}, \ and\
  \bibinfo {author} {\bibfnamefont {A.~A.}\ \bibnamefont {Mohamad}},\
  }\bibfield  {title} {\enquote {\bibinfo {title} {Sedimentation of elliptical
  particles using immersed boundary --- lattice {Boltzmann} method: A
  complementary repulsive force model},}\ }\href {\doibase
  10.1016/j.molliq.2018.04.075} {\bibfield  {journal} {\bibinfo  {journal} {J.
  Mol. Liq.}\ }\textbf {\bibinfo {volume} {262}},\ \bibinfo {pages} {180--193}
  (\bibinfo {year} {2018})}\BibitemShut {NoStop}%
\bibitem [{\citenamefont {Singh}\ \emph {et~al.}(2018)\citenamefont {Singh},
  \citenamefont {Gompper},\ and\ \citenamefont {Winkler}}]{Singh2018}%
  \BibitemOpen
  \bibfield  {author} {\bibinfo {author} {\bibfnamefont {Sunil~P.}\
  \bibnamefont {Singh}}, \bibinfo {author} {\bibfnamefont {Gerhard}\
  \bibnamefont {Gompper}}, \ and\ \bibinfo {author} {\bibfnamefont {Roland~G.}\
  \bibnamefont {Winkler}},\ }\bibfield  {title} {\enquote {\bibinfo {title}
  {Steady state sedimentation of ultrasoft colloids},}\ }\href {\doibase
  10.1063/1.5001886} {\bibfield  {journal} {\bibinfo  {journal} {J. Chem.
  Phys.}\ }\textbf {\bibinfo {volume} {148}},\ \bibinfo {pages} {084901}
  (\bibinfo {year} {2018})}\BibitemShut {NoStop}%
\bibitem [{\citenamefont {Hu}\ \emph {et~al.}(2010)\citenamefont {Hu},
  \citenamefont {Hecht},\ and\ \citenamefont {Gruner}}]{Hu2010}%
  \BibitemOpen
  \bibfield  {author} {\bibinfo {author} {\bibfnamefont {Liangbing}\
  \bibnamefont {Hu}}, \bibinfo {author} {\bibfnamefont {David~S.}\ \bibnamefont
  {Hecht}}, \ and\ \bibinfo {author} {\bibfnamefont {George}\ \bibnamefont
  {Gruner}},\ }\bibfield  {title} {\enquote {\bibinfo {title} {Carbon nanotube
  thin films: fabrication, properties, and applications},}\ }\href {\doibase
  10.1021/cr9002962} {\bibfield  {journal} {\bibinfo  {journal} {Chem. Rev.}\
  }\textbf {\bibinfo {volume} {110}},\ \bibinfo {pages} {5790--5844} (\bibinfo
  {year} {2010})}\BibitemShut {NoStop}%
\bibitem [{\citenamefont {Mohraz}\ and\ \citenamefont
  {Solomon}(2005)}]{Mohraz2005}%
  \BibitemOpen
  \bibfield  {author} {\bibinfo {author} {\bibfnamefont {Ali}\ \bibnamefont
  {Mohraz}}\ and\ \bibinfo {author} {\bibfnamefont {Michael~J.}\ \bibnamefont
  {Solomon}},\ }\bibfield  {title} {\enquote {\bibinfo {title} {Direct
  visualization of colloidal rod assembly by confocal microscopy},}\ }\href
  {\doibase 10.1021/la046908a} {\bibfield  {journal} {\bibinfo  {journal}
  {Langmuir}\ }\textbf {\bibinfo {volume} {21}},\ \bibinfo {pages} {5298--5306}
  (\bibinfo {year} {2005})}\BibitemShut {NoStop}%
\bibitem [{\citenamefont {Onsager}(1949)}]{Onsager1949}%
  \BibitemOpen
  \bibfield  {author} {\bibinfo {author} {\bibfnamefont {Lars}\ \bibnamefont
  {Onsager}},\ }\bibfield  {title} {\enquote {\bibinfo {title} {The effects of
  shape on the interaction of colloidal particles},}\ }\href {\doibase
  10.1111/j.1749-6632.1949.tb27296.x} {\bibfield  {journal} {\bibinfo
  {journal} {Ann. N. Y. Acad. Sci.}\ }\textbf {\bibinfo {volume} {51}},\
  \bibinfo {pages} {627--659} (\bibinfo {year} {1949})}\BibitemShut {NoStop}%
\bibitem [{\citenamefont {Adams}\ \emph {et~al.}(1998)\citenamefont {Adams},
  \citenamefont {Dogic}, \citenamefont {Keller},\ and\ \citenamefont
  {Fraden}}]{Adams1998}%
  \BibitemOpen
  \bibfield  {author} {\bibinfo {author} {\bibfnamefont {Marie}\ \bibnamefont
  {Adams}}, \bibinfo {author} {\bibfnamefont {Zvonimir}\ \bibnamefont {Dogic}},
  \bibinfo {author} {\bibfnamefont {Sarah~L.}\ \bibnamefont {Keller}}, \ and\
  \bibinfo {author} {\bibfnamefont {Seth}\ \bibnamefont {Fraden}},\ }\bibfield
  {title} {\enquote {\bibinfo {title} {Entropically driven microphase
  transitions in mixtures of colloidal rods and spheres},}\ }\href {\doibase
  10.1038/30700} {\bibfield  {journal} {\bibinfo  {journal} {Nature}\ }\textbf
  {\bibinfo {volume} {393}},\ \bibinfo {pages} {349--352} (\bibinfo {year}
  {1998})}\BibitemShut {NoStop}%
\bibitem [{\citenamefont {Maeda}\ and\ \citenamefont
  {Maeda}(2003)}]{Maeda2003}%
  \BibitemOpen
  \bibfield  {author} {\bibinfo {author} {\bibfnamefont {Hideatsu}\
  \bibnamefont {Maeda}}\ and\ \bibinfo {author} {\bibfnamefont {Yoshiko}\
  \bibnamefont {Maeda}},\ }\bibfield  {title} {\enquote {\bibinfo {title}
  {Liquid crystal formation in suspensions of hard rodlike colloidal particles:
  direct observation of particle arrangement and self-ordering behavior},}\
  }\href {\doibase 10.1103/PhysRevLett.90.018303} {\bibfield  {journal}
  {\bibinfo  {journal} {Phys. Rev. Lett.}\ }\textbf {\bibinfo {volume} {90}},\
  \bibinfo {pages} {018303} (\bibinfo {year} {2003})}\BibitemShut {NoStop}%
\bibitem [{\citenamefont {Alargova}\ \emph {et~al.}(2004)\citenamefont
  {Alargova}, \citenamefont {Bhatt}, \citenamefont {Paunov},\ and\
  \citenamefont {Velev}}]{Alargova2004}%
  \BibitemOpen
  \bibfield  {author} {\bibinfo {author} {\bibfnamefont {Rossitza~G.}\
  \bibnamefont {Alargova}}, \bibinfo {author} {\bibfnamefont {Ketan~H.}\
  \bibnamefont {Bhatt}}, \bibinfo {author} {\bibfnamefont {Vesselin~N.}\
  \bibnamefont {Paunov}}, \ and\ \bibinfo {author} {\bibfnamefont {Orlin~D.}\
  \bibnamefont {Velev}},\ }\bibfield  {title} {\enquote {\bibinfo {title}
  {Scalable synthesis of a new class of polymer microrods by a liquid--liquid
  dispersion technique},}\ }\href {\doibase 10.1002/adma.200400112} {\bibfield
  {journal} {\bibinfo  {journal} {Adv. Mater.}\ }\textbf {\bibinfo {volume}
  {16}},\ \bibinfo {pages} {1653--1657} (\bibinfo {year} {2004})}\BibitemShut
  {NoStop}%
\bibitem [{\citenamefont {Kayser}\ and\ \citenamefont
  {Ravech\'e}(1978)}]{Kayser1978}%
  \BibitemOpen
  \bibfield  {author} {\bibinfo {author} {\bibfnamefont {Richard~F.}\
  \bibnamefont {Kayser}}\ and\ \bibinfo {author} {\bibfnamefont {Harold~J.}\
  \bibnamefont {Ravech\'e}},\ }\bibfield  {title} {\enquote {\bibinfo {title}
  {Bifurcation in {Onsager}'s model of the isotropic-nematic transition},}\
  }\href {\doibase 10.1103/PhysRevA.17.2067} {\bibfield  {journal} {\bibinfo
  {journal} {Phys. Rev. A}\ }\textbf {\bibinfo {volume} {17}},\ \bibinfo
  {pages} {2067--2072} (\bibinfo {year} {1978})}\BibitemShut {NoStop}%
\bibitem [{\citenamefont {Frenkel}\ and\ \citenamefont
  {Eppenga}(1985)}]{Frenkel1985}%
  \BibitemOpen
  \bibfield  {author} {\bibinfo {author} {\bibfnamefont {D.}~\bibnamefont
  {Frenkel}}\ and\ \bibinfo {author} {\bibfnamefont {R.}~\bibnamefont
  {Eppenga}},\ }\bibfield  {title} {\enquote {\bibinfo {title} {Evidence for
  algebraic orientational order in a two-dimensional hard-core nematic},}\
  }\href {\doibase 10.1103/PhysRevA.31.1776} {\bibfield  {journal} {\bibinfo
  {journal} {Phys. Rev. A}\ }\textbf {\bibinfo {volume} {31}},\ \bibinfo
  {pages} {1776} (\bibinfo {year} {1985})}\BibitemShut {NoStop}%
\bibitem [{\citenamefont {Bates}\ and\ \citenamefont
  {Frenkel}(2000)}]{Bates2000}%
  \BibitemOpen
  \bibfield  {author} {\bibinfo {author} {\bibfnamefont {Martin~A.}\
  \bibnamefont {Bates}}\ and\ \bibinfo {author} {\bibfnamefont {Daan}\
  \bibnamefont {Frenkel}},\ }\bibfield  {title} {\enquote {\bibinfo {title}
  {Phase behavior of two-dimensional hard rod fluids},}\ }\href {\doibase
  10.1063/1.481637} {\bibfield  {journal} {\bibinfo  {journal} {J. Chem.
  Phys.}\ }\textbf {\bibinfo {volume} {112}},\ \bibinfo {pages} {10034--10041}
  (\bibinfo {year} {2000})}\BibitemShut {NoStop}%
\bibitem [{\citenamefont {Ghosh}\ and\ \citenamefont {Dhar}(2007)}]{Ghosh2007}%
  \BibitemOpen
  \bibfield  {author} {\bibinfo {author} {\bibfnamefont {Anandamohan}\
  \bibnamefont {Ghosh}}\ and\ \bibinfo {author} {\bibfnamefont {Deepak}\
  \bibnamefont {Dhar}},\ }\bibfield  {title} {\enquote {\bibinfo {title} {On
  the orientational ordering of long rods on a lattice},}\ }\href {\doibase
  10.1209/0295-5075/78/20003} {\bibfield  {journal} {\bibinfo  {journal} {EPL
  (Europhys. Lett.)}\ }\textbf {\bibinfo {volume} {78}},\ \bibinfo {pages}
  {20003} (\bibinfo {year} {2007})}\BibitemShut {NoStop}%
\bibitem [{\citenamefont {L\'{o}pez}\ \emph {et~al.}(2010)\citenamefont
  {L\'{o}pez}, \citenamefont {Linares}, \citenamefont {Ramirez-Pastor},\ and\
  \citenamefont {Cannas}}]{Lopez2010}%
  \BibitemOpen
  \bibfield  {author} {\bibinfo {author} {\bibfnamefont {L.~G.}\ \bibnamefont
  {L\'{o}pez}}, \bibinfo {author} {\bibfnamefont {D.~H.}\ \bibnamefont
  {Linares}}, \bibinfo {author} {\bibfnamefont {A.~J.}\ \bibnamefont
  {Ramirez-Pastor}}, \ and\ \bibinfo {author} {\bibfnamefont {S.~A.}\
  \bibnamefont {Cannas}},\ }\bibfield  {title} {\enquote {\bibinfo {title}
  {Phase diagram of self-assembled rigid rods on two-dimensional lattices:
  Theory and {M}onte {C}arlo simulations},}\ }\href {\doibase
  10.1063/1.3496482} {\bibfield  {journal} {\bibinfo  {journal} {J. Chem.
  Phys.}\ }\textbf {\bibinfo {volume} {133}},\ \bibinfo {pages} {134706}
  (\bibinfo {year} {2010})}\BibitemShut {NoStop}%
\bibitem [{\citenamefont {Lon{\v{c}}arevi{\'{c}}}\ \emph
  {et~al.}(2010)\citenamefont {Lon{\v{c}}arevi{\'{c}}}, \citenamefont
  {Jak{\v{s}}i{\'{c}}}, \citenamefont {Vrhovac},\ and\ \citenamefont
  {Budinski-Petkovi{\'{c}}}}]{Loncarevic2010}%
  \BibitemOpen
  \bibfield  {author} {\bibinfo {author} {\bibfnamefont {I.}~\bibnamefont
  {Lon{\v{c}}arevi{\'{c}}}}, \bibinfo {author} {\bibfnamefont {M.~Z.}\
  \bibnamefont {Jak{\v{s}}i{\'{c}}}}, \bibinfo {author} {\bibfnamefont {B.~S.}\
  \bibnamefont {Vrhovac}}, \ and\ \bibinfo {author} {\bibfnamefont {Lj.}\
  \bibnamefont {Budinski-Petkovi{\'{c}}}},\ }\bibfield  {title} {\enquote
  {\bibinfo {title} {Irreversible deposition of extended objects with
  diffusional relaxation on discrete substrates},}\ }\href {\doibase
  10.1140/epjb/e2010-00010-1} {\bibfield  {journal} {\bibinfo  {journal} {Eur.
  Phys. J. B}\ }\textbf {\bibinfo {volume} {73}},\ \bibinfo {pages} {439--445}
  (\bibinfo {year} {2010})}\BibitemShut {NoStop}%
\bibitem [{\citenamefont {Matoz-Fernandez}\ \emph {et~al.}(2012)\citenamefont
  {Matoz-Fernandez}, \citenamefont {Linares},\ and\ \citenamefont
  {Ramirez-Pastor}}]{Matoz-Fernandez2012}%
  \BibitemOpen
  \bibfield  {author} {\bibinfo {author} {\bibfnamefont {D.~A.}\ \bibnamefont
  {Matoz-Fernandez}}, \bibinfo {author} {\bibfnamefont {D.~H.}\ \bibnamefont
  {Linares}}, \ and\ \bibinfo {author} {\bibfnamefont {A.~J.}\ \bibnamefont
  {Ramirez-Pastor}},\ }\bibfield  {title} {\enquote {\bibinfo {title}
  {Nonmonotonic size dependence of the critical concentration in {2D}
  percolation of straight rigid rods under equilibrium conditions},}\ }\href
  {\doibase 10.1140/epjb/e2012-30389-2} {\bibfield  {journal} {\bibinfo
  {journal} {Eur. Phys. J. B}\ }\textbf {\bibinfo {volume} {85}},\ \bibinfo
  {pages} {1--7} (\bibinfo {year} {2012})}\BibitemShut {NoStop}%
\bibitem [{\citenamefont {Kundu}\ and\ \citenamefont
  {Rajesh}(2013)}]{Kundu2013}%
  \BibitemOpen
  \bibfield  {author} {\bibinfo {author} {\bibfnamefont {Joyjit}\ \bibnamefont
  {Kundu}}\ and\ \bibinfo {author} {\bibfnamefont {R.}~\bibnamefont {Rajesh}},\
  }\bibfield  {title} {\enquote {\bibinfo {title} {Reentrant disordered phase
  in a system of repulsive rods on a {Bethe}-like lattice},}\ }\href {\doibase
  10.1103/PhysRevE.88.012134} {\bibfield  {journal} {\bibinfo  {journal} {Phys.
  Rev. E}\ }\textbf {\bibinfo {volume} {88}},\ \bibinfo {pages} {012134}
  (\bibinfo {year} {2013})}\BibitemShut {NoStop}%
\bibitem [{\citenamefont {Kundu}\ \emph {et~al.}(2013)\citenamefont {Kundu},
  \citenamefont {Rajesh}, \citenamefont {Dhar},\ and\ \citenamefont
  {Stilck}}]{Kundu2013a}%
  \BibitemOpen
  \bibfield  {author} {\bibinfo {author} {\bibfnamefont {Joyjit}\ \bibnamefont
  {Kundu}}, \bibinfo {author} {\bibfnamefont {R.}~\bibnamefont {Rajesh}},
  \bibinfo {author} {\bibfnamefont {Deepak}\ \bibnamefont {Dhar}}, \ and\
  \bibinfo {author} {\bibfnamefont {J\"urgen~F.}\ \bibnamefont {Stilck}},\
  }\bibfield  {title} {\enquote {\bibinfo {title} {Nematic-disordered phase
  transition in systems of long rigid rods on two-dimensional lattices},}\
  }\href {\doibase 10.1103/PhysRevE.87.032103} {\bibfield  {journal} {\bibinfo
  {journal} {Phys. Rev. E}\ }\textbf {\bibinfo {volume} {87}},\ \bibinfo
  {pages} {032103} (\bibinfo {year} {2013})}\BibitemShut {NoStop}%
\bibitem [{\citenamefont {Lebovka}\ \emph {et~al.}(2017)\citenamefont
  {Lebovka}, \citenamefont {Tarasevich}, \citenamefont {Gigiberiya},\ and\
  \citenamefont {Vygornitskii}}]{Lebovka2017}%
  \BibitemOpen
  \bibfield  {author} {\bibinfo {author} {\bibfnamefont {N.~I.}\ \bibnamefont
  {Lebovka}}, \bibinfo {author} {\bibfnamefont {Yu.~Yu.}\ \bibnamefont
  {Tarasevich}}, \bibinfo {author} {\bibfnamefont {V.~A.}\ \bibnamefont
  {Gigiberiya}}, \ and\ \bibinfo {author} {\bibfnamefont {N.~V.}\ \bibnamefont
  {Vygornitskii}},\ }\bibfield  {title} {\enquote {\bibinfo {title}
  {Diffusion-driven self-assembly of rodlike particles: {M}onte {C}arlo
  simulation on a square lattice},}\ }\href {\doibase
  10.1103/PhysRevE.95.052130} {\bibfield  {journal} {\bibinfo  {journal} {Phys.
  Rev. E}\ }\textbf {\bibinfo {volume} {95}},\ \bibinfo {pages} {052130}
  (\bibinfo {year} {2017})}\BibitemShut {NoStop}%
\bibitem [{\citenamefont {Vold}(1959)}]{Vold1959}%
  \BibitemOpen
  \bibfield  {author} {\bibinfo {author} {\bibfnamefont {Marjorie~J.}\
  \bibnamefont {Vold}},\ }\bibfield  {title} {\enquote {\bibinfo {title}
  {Sediment volume and structure in dispersions of anisometric particles},}\
  }\href {\doibase 10.1021/j150580a011} {\bibfield  {journal} {\bibinfo
  {journal} {J. Phys. Chem.}\ }\textbf {\bibinfo {volume} {63}},\ \bibinfo
  {pages} {1608--1612} (\bibinfo {year} {1959})}\BibitemShut {NoStop}%
\bibitem [{\citenamefont {Biben}\ \emph {et~al.}(1993)\citenamefont {Biben},
  \citenamefont {Hansen},\ and\ \citenamefont {Barrat}}]{Biben1993}%
  \BibitemOpen
  \bibfield  {author} {\bibinfo {author} {\bibfnamefont {Thierry}\ \bibnamefont
  {Biben}}, \bibinfo {author} {\bibfnamefont {Jean-Pierre}\ \bibnamefont
  {Hansen}}, \ and\ \bibinfo {author} {\bibfnamefont {Jean-Louis}\ \bibnamefont
  {Barrat}},\ }\bibfield  {title} {\enquote {\bibinfo {title} {Density profiles
  of concentrated colloidal suspensions in sedimentation equilibrium},}\ }\href
  {\doibase 10.1063/1.464726} {\bibfield  {journal} {\bibinfo  {journal} {J.
  Chem. Phys.}\ }\textbf {\bibinfo {volume} {98}},\ \bibinfo {pages}
  {7330--7344} (\bibinfo {year} {1993})}\BibitemShut {NoStop}%
\bibitem [{\citenamefont {Guazzelli}\ and\ \citenamefont
  {Hinch}(2011)}]{Guazzelli2011}%
  \BibitemOpen
  \bibfield  {author} {\bibinfo {author} {\bibfnamefont {\'{E}lisabeth}\
  \bibnamefont {Guazzelli}}\ and\ \bibinfo {author} {\bibfnamefont {John}\
  \bibnamefont {Hinch}},\ }\bibfield  {title} {\enquote {\bibinfo {title}
  {Fluctuations and instability in sedimentation},}\ }\href {\doibase
  10.1146/annurev-fluid-122109-160736} {\bibfield  {journal} {\bibinfo
  {journal} {Annu. Rev. Fluid Mech.}\ }\textbf {\bibinfo {volume} {43}},\
  \bibinfo {pages} {97--116} (\bibinfo {year} {2011})}\BibitemShut {NoStop}%
\bibitem [{\citenamefont {Koch}\ and\ \citenamefont
  {Shaqfeh}(1989)}]{Koch1989}%
  \BibitemOpen
  \bibfield  {author} {\bibinfo {author} {\bibfnamefont {Donald~L.}\
  \bibnamefont {Koch}}\ and\ \bibinfo {author} {\bibfnamefont {Eric S.~G.}\
  \bibnamefont {Shaqfeh}},\ }\bibfield  {title} {\enquote {\bibinfo {title}
  {The instability of a dispersion of sedimenting spheroids},}\ }\href
  {\doibase 10.1017/S0022112089003204} {\bibfield  {journal} {\bibinfo
  {journal} {J. Fluid Mech.}\ }\textbf {\bibinfo {volume} {209}},\ \bibinfo
  {pages} {521--542} (\bibinfo {year} {1989})}\BibitemShut {NoStop}%
\bibitem [{\citenamefont {Herzhaft}\ \emph {et~al.}(1996)\citenamefont
  {Herzhaft}, \citenamefont {Guazzelli}, \citenamefont {Mackaplow},\ and\
  \citenamefont {Shaqfeh}}]{Herzhaft1996}%
  \BibitemOpen
  \bibfield  {author} {\bibinfo {author} {\bibfnamefont {Benjamin}\
  \bibnamefont {Herzhaft}}, \bibinfo {author} {\bibfnamefont {\'{E}lisabeth}\
  \bibnamefont {Guazzelli}}, \bibinfo {author} {\bibfnamefont {Michael~B.}\
  \bibnamefont {Mackaplow}}, \ and\ \bibinfo {author} {\bibfnamefont {Eric
  S.~G.}\ \bibnamefont {Shaqfeh}},\ }\bibfield  {title} {\enquote {\bibinfo
  {title} {Experimental investigation of the sedimentation of a dilute fiber
  suspension},}\ }\href {\doibase 10.1103/PhysRevLett.77.290} {\bibfield
  {journal} {\bibinfo  {journal} {Phys. Rev. Lett.}\ }\textbf {\bibinfo
  {volume} {77}},\ \bibinfo {pages} {290--293} (\bibinfo {year}
  {1996})}\BibitemShut {NoStop}%
\bibitem [{\citenamefont {Herzhaft}\ and\ \citenamefont
  {Guazzelli}(1999)}]{Herzhaft1999}%
  \BibitemOpen
  \bibfield  {author} {\bibinfo {author} {\bibfnamefont {Benjamin}\
  \bibnamefont {Herzhaft}}\ and\ \bibinfo {author} {\bibfnamefont
  {\'{E}lisabeth}\ \bibnamefont {Guazzelli}},\ }\bibfield  {title} {\enquote
  {\bibinfo {title} {Experimental study of the sedimentation of dilute and
  semi-dilute suspensions of fibres},}\ }\href {\doibase
  10.1017/S0022112099004152} {\bibfield  {journal} {\bibinfo  {journal} {J.
  Fluid Mech.}\ }\textbf {\bibinfo {volume} {384}},\ \bibinfo {pages}
  {133--158} (\bibinfo {year} {1999})}\BibitemShut {NoStop}%
\bibitem [{\citenamefont {Metzger}\ \emph {et~al.}(2005)\citenamefont
  {Metzger}, \citenamefont {Guazzelli},\ and\ \citenamefont
  {Butler}}]{Metzger2005}%
  \BibitemOpen
  \bibfield  {author} {\bibinfo {author} {\bibfnamefont {Bloen}\ \bibnamefont
  {Metzger}}, \bibinfo {author} {\bibfnamefont {\'Elisabeth}\ \bibnamefont
  {Guazzelli}}, \ and\ \bibinfo {author} {\bibfnamefont {Jason~E.}\
  \bibnamefont {Butler}},\ }\bibfield  {title} {\enquote {\bibinfo {title}
  {Large-scale streamers in the sedimentation of a dilute fiber suspension},}\
  }\href {\doibase 10.1103/PhysRevLett.95.164506} {\bibfield  {journal}
  {\bibinfo  {journal} {Phys. Rev. Lett.}\ }\textbf {\bibinfo {volume} {95}},\
  \bibinfo {pages} {164506} (\bibinfo {year} {2005})}\BibitemShut {NoStop}%
\bibitem [{\citenamefont {Metzger}\ \emph {et~al.}(2007)\citenamefont
  {Metzger}, \citenamefont {Butler},\ and\ \citenamefont
  {Guazzelli}}]{Metzger2007}%
  \BibitemOpen
  \bibfield  {author} {\bibinfo {author} {\bibfnamefont {Bloen}\ \bibnamefont
  {Metzger}}, \bibinfo {author} {\bibfnamefont {Jason~E.}\ \bibnamefont
  {Butler}}, \ and\ \bibinfo {author} {\bibfnamefont {\'{E}lisabeth}\
  \bibnamefont {Guazzelli}},\ }\bibfield  {title} {\enquote {\bibinfo {title}
  {Experimental investigation of the instability of a sedimenting suspension of
  fibres},}\ }\href {\doibase 10.1017/S0022112006004460} {\bibfield  {journal}
  {\bibinfo  {journal} {J. Fluid Mech.}\ }\textbf {\bibinfo {volume} {575}},\
  \bibinfo {pages} {307--332} (\bibinfo {year} {2007})}\BibitemShut {NoStop}%
\bibitem [{\citenamefont {Mackaplow}\ and\ \citenamefont
  {Shaqfeh}(1998)}]{Mackaplow1998}%
  \BibitemOpen
  \bibfield  {author} {\bibinfo {author} {\bibfnamefont {Michael~B.}\
  \bibnamefont {Mackaplow}}\ and\ \bibinfo {author} {\bibfnamefont {Eric
  S.~G.}\ \bibnamefont {Shaqfeh}},\ }\bibfield  {title} {\enquote {\bibinfo
  {title} {A numerical study of the sedimentation of fibre suspensions},}\
  }\href {\doibase 10.1017/S0022112098002663} {\bibfield  {journal} {\bibinfo
  {journal} {J. Fluid Mech.}\ }\textbf {\bibinfo {volume} {376}},\ \bibinfo
  {pages} {149--182} (\bibinfo {year} {1998})}\BibitemShut {NoStop}%
\bibitem [{\citenamefont {Butler}\ and\ \citenamefont
  {Shaqfeh}(2002)}]{Butler2002}%
  \BibitemOpen
  \bibfield  {author} {\bibinfo {author} {\bibfnamefont {Jason~E.}\
  \bibnamefont {Butler}}\ and\ \bibinfo {author} {\bibfnamefont {Eric S.~G.}\
  \bibnamefont {Shaqfeh}},\ }\bibfield  {title} {\enquote {\bibinfo {title}
  {Dynamic simulations of the inhomogeneous sedimentation of rigid fibres},}\
  }\href {\doibase 10.1017/S0022112002001544} {\bibfield  {journal} {\bibinfo
  {journal} {J. Fluid Mech.}\ }\textbf {\bibinfo {volume} {468}},\ \bibinfo
  {pages} {205--237} (\bibinfo {year} {2002})}\BibitemShut {NoStop}%
\bibitem [{\citenamefont {Saintillan}\ \emph {et~al.}(2005)\citenamefont
  {Saintillan}, \citenamefont {Darve},\ and\ \citenamefont
  {Shaqfeh}}]{Saintillan2005}%
  \BibitemOpen
  \bibfield  {author} {\bibinfo {author} {\bibfnamefont {David}\ \bibnamefont
  {Saintillan}}, \bibinfo {author} {\bibfnamefont {Eric}\ \bibnamefont
  {Darve}}, \ and\ \bibinfo {author} {\bibfnamefont {Eric S.~G.}\ \bibnamefont
  {Shaqfeh}},\ }\bibfield  {title} {\enquote {\bibinfo {title} {A smooth
  particle-mesh {Ewald} algorithm for {Stokes} suspension simulations: The
  sedimentation of fibers},}\ }\href {\doibase 10.1063/1.1862262} {\bibfield
  {journal} {\bibinfo  {journal} {Phys. Fluids}\ }\textbf {\bibinfo {volume}
  {17}},\ \bibinfo {pages} {033301} (\bibinfo {year} {2005})}\BibitemShut
  {NoStop}%
\bibitem [{\citenamefont {Helzel}\ and\ \citenamefont
  {Tzavaras}(2017)}]{Helzel2017}%
  \BibitemOpen
  \bibfield  {author} {\bibinfo {author} {\bibfnamefont {C.}~\bibnamefont
  {Helzel}}\ and\ \bibinfo {author} {\bibfnamefont {A.}~\bibnamefont
  {Tzavaras}},\ }\bibfield  {title} {\enquote {\bibinfo {title} {A kinetic
  model for the sedimentation of rod--like particles},}\ }\href {\doibase
  10.1137/15M1023907} {\bibfield  {journal} {\bibinfo  {journal} {Multiscale
  Modeling \& Simulation}\ }\textbf {\bibinfo {volume} {15}},\ \bibinfo {pages}
  {500--536} (\bibinfo {year} {2017})}\BibitemShut {NoStop}%
\bibitem [{\citenamefont {Kuusela}\ \emph {et~al.}(2001)\citenamefont
  {Kuusela}, \citenamefont {H{\"o}fler},\ and\ \citenamefont
  {Schwarzer}}]{Kuusela2001}%
  \BibitemOpen
  \bibfield  {author} {\bibinfo {author} {\bibfnamefont {Esa}\ \bibnamefont
  {Kuusela}}, \bibinfo {author} {\bibfnamefont {Kai}\ \bibnamefont
  {H{\"o}fler}}, \ and\ \bibinfo {author} {\bibfnamefont {Stefan}\ \bibnamefont
  {Schwarzer}},\ }\bibfield  {title} {\enquote {\bibinfo {title} {Computation
  of particle settling speed and orientation distribution in suspensions of
  prolate spheroids},}\ }\href {\doibase 10.1023/A:1011900103361} {\bibfield
  {journal} {\bibinfo  {journal} {J. Eng. Math.}\ }\textbf {\bibinfo {volume}
  {41}},\ \bibinfo {pages} {221--235} (\bibinfo {year} {2001})}\BibitemShut
  {NoStop}%
\bibitem [{\citenamefont {Kuusela}\ \emph {et~al.}(2003)\citenamefont
  {Kuusela}, \citenamefont {Lahtinen},\ and\ \citenamefont
  {Ala-Nissila}}]{Kuusela2003}%
  \BibitemOpen
  \bibfield  {author} {\bibinfo {author} {\bibfnamefont {E.}~\bibnamefont
  {Kuusela}}, \bibinfo {author} {\bibfnamefont {J.~M.}\ \bibnamefont
  {Lahtinen}}, \ and\ \bibinfo {author} {\bibfnamefont {T.}~\bibnamefont
  {Ala-Nissila}},\ }\bibfield  {title} {\enquote {\bibinfo {title} {Collective
  effects in settling of spheroids under steady-state sedimentation},}\ }\href
  {\doibase 10.1103/PhysRevLett.90.094502} {\bibfield  {journal} {\bibinfo
  {journal} {Phys. Rev. Lett.}\ }\textbf {\bibinfo {volume} {90}},\ \bibinfo
  {pages} {094502} (\bibinfo {year} {2003})}\BibitemShut {NoStop}%
\bibitem [{\citenamefont {Salmela}\ \emph {et~al.}(2007)\citenamefont
  {Salmela}, \citenamefont {Martinez},\ and\ \citenamefont
  {Kataja}}]{Salmela2007}%
  \BibitemOpen
  \bibfield  {author} {\bibinfo {author} {\bibfnamefont {J.}~\bibnamefont
  {Salmela}}, \bibinfo {author} {\bibfnamefont {D.~M.}\ \bibnamefont
  {Martinez}}, \ and\ \bibinfo {author} {\bibfnamefont {M.}~\bibnamefont
  {Kataja}},\ }\bibfield  {title} {\enquote {\bibinfo {title} {Settling of
  dilute and semidilute fiber suspensions at finite {Re}},}\ }\href {\doibase
  10.1002/aic.11245} {\bibfield  {journal} {\bibinfo  {journal} {{AIChE} J.}\
  }\textbf {\bibinfo {volume} {53}},\ \bibinfo {pages} {1916--1923} (\bibinfo
  {year} {2007})}\BibitemShut {NoStop}%
\bibitem [{\citenamefont {Wang}\ and\ \citenamefont {Layton}(2009)}]{Wang2009}%
  \BibitemOpen
  \bibfield  {author} {\bibinfo {author} {\bibfnamefont {Jin}\ \bibnamefont
  {Wang}}\ and\ \bibinfo {author} {\bibfnamefont {Anita}\ \bibnamefont
  {Layton}},\ }\bibfield  {title} {\enquote {\bibinfo {title} {Numerical
  simulations of fiber sedimentation in {Navier}---{Stokes} flows},}\ }\href
  {http://www.global-sci.org/v1/cicp/freedownload/v5_61.pdf} {\bibfield
  {journal} {\bibinfo  {journal} {Commun. Comput. Phys}\ }\textbf {\bibinfo
  {volume} {5}},\ \bibinfo {pages} {61--83} (\bibinfo {year}
  {2009})}\BibitemShut {NoStop}%
\bibitem [{\citenamefont {Gustavsson}\ and\ \citenamefont
  {Tornberg}(2009)}]{Gustavsson2009}%
  \BibitemOpen
  \bibfield  {author} {\bibinfo {author} {\bibfnamefont {K.}~\bibnamefont
  {Gustavsson}}\ and\ \bibinfo {author} {\bibfnamefont {A.-K.}\ \bibnamefont
  {Tornberg}},\ }\bibfield  {title} {\enquote {\bibinfo {title} {Gravity
  induced sedimentation of slender fibers},}\ }\href {\doibase
  10.1063/1.3273091} {\bibfield  {journal} {\bibinfo  {journal} {Phys. Fluids}\
  }\textbf {\bibinfo {volume} {21}},\ \bibinfo {pages} {123301} (\bibinfo
  {year} {2009})}\BibitemShut {NoStop}%
\bibitem [{\citenamefont {Vroege}\ and\ \citenamefont
  {Lekkerkerker}(1992)}]{Vroege1992}%
  \BibitemOpen
  \bibfield  {author} {\bibinfo {author} {\bibfnamefont {G.~J.}\ \bibnamefont
  {Vroege}}\ and\ \bibinfo {author} {\bibfnamefont {H.~N.~W.}\ \bibnamefont
  {Lekkerkerker}},\ }\bibfield  {title} {\enquote {\bibinfo {title} {Phase
  transitions in lyotropic colloidal and polymer liquid crystals},}\ }\href
  {\doibase 10.1088/0034-4885/55/8/003} {\bibfield  {journal} {\bibinfo
  {journal} {Rep. Prog. Phys.}\ }\textbf {\bibinfo {volume} {55}},\ \bibinfo
  {pages} {1241--1309} (\bibinfo {year} {1992})}\BibitemShut {NoStop}%
\bibitem [{\citenamefont {Philipse}(1997)}]{Philipse1997}%
  \BibitemOpen
  \bibfield  {author} {\bibinfo {author} {\bibfnamefont {Albert~P.}\
  \bibnamefont {Philipse}},\ }\bibfield  {title} {\enquote {\bibinfo {title}
  {Colloidal sedimentation (and filtration)},}\ }\href {\doibase
  10.1016/S1359-0294(97)80027-1} {\bibfield  {journal} {\bibinfo  {journal}
  {Curr. Opin. Colloid Interface Sci.}\ }\textbf {\bibinfo {volume} {2}},\
  \bibinfo {pages} {200--206} (\bibinfo {year} {1997})}\BibitemShut {NoStop}%
\bibitem [{\citenamefont {Allen}\ \emph {et~al.}(1999)\citenamefont {Allen},
  \citenamefont {Goulding},\ and\ \citenamefont {Hansen}}]{Allen1999}%
  \BibitemOpen
  \bibfield  {author} {\bibinfo {author} {\bibfnamefont {Rosalind}\
  \bibnamefont {Allen}}, \bibinfo {author} {\bibfnamefont {David}\ \bibnamefont
  {Goulding}}, \ and\ \bibinfo {author} {\bibfnamefont {Jean-Pierre}\
  \bibnamefont {Hansen}},\ }\bibfield  {title} {\enquote {\bibinfo {title}
  {Sedimentation equilibria of colloidal hard rod dispersions},}\ }\href
  {\doibase 10.1039/A905089B} {\bibfield  {journal} {\bibinfo  {journal}
  {PhysChemComm}\ }\textbf {\bibinfo {volume} {2}},\ \bibinfo {pages} {30--33}
  (\bibinfo {year} {1999})}\BibitemShut {NoStop}%
\bibitem [{\citenamefont {Savenko}\ and\ \citenamefont
  {Dijkstra}(2004)}]{Savenko2004}%
  \BibitemOpen
  \bibfield  {author} {\bibinfo {author} {\bibfnamefont {S.~V.}\ \bibnamefont
  {Savenko}}\ and\ \bibinfo {author} {\bibfnamefont {Marjolein}\ \bibnamefont
  {Dijkstra}},\ }\bibfield  {title} {\enquote {\bibinfo {title} {Sedimentation
  and multiphase equilibria in suspensions of colloidal hard rods},}\ }\href
  {\doibase 10.1103/PhysRevE.70.051401} {\bibfield  {journal} {\bibinfo
  {journal} {Phys. Rev. E}\ }\textbf {\bibinfo {volume} {70}},\ \bibinfo
  {pages} {051401} (\bibinfo {year} {2004})}\BibitemShut {NoStop}%
\bibitem [{\citenamefont {Sanz}\ and\ \citenamefont
  {Marenduzzo}(2010)}]{Sanz2010}%
  \BibitemOpen
  \bibfield  {author} {\bibinfo {author} {\bibfnamefont {E.}~\bibnamefont
  {Sanz}}\ and\ \bibinfo {author} {\bibfnamefont {D.}~\bibnamefont
  {Marenduzzo}},\ }\bibfield  {title} {\enquote {\bibinfo {title} {Dynamic
  {Monte} {Carlo} versus {Brownian} dynamics: A comparison for self-diffusion
  and crystallization in colloidal fluids},}\ }\href {\doibase
  10.1063/1.3414827} {\bibfield  {journal} {\bibinfo  {journal} {J. Chem.
  Phys.}\ }\textbf {\bibinfo {volume} {132}},\ \bibinfo {pages} {194102}
  (\bibinfo {year} {2010})}\BibitemShut {NoStop}%
\bibitem [{\citenamefont {Romano}\ \emph {et~al.}(2011)\citenamefont {Romano},
  \citenamefont {De~Michele}, \citenamefont {Marenduzzo},\ and\ \citenamefont
  {Sanz}}]{Romano2011}%
  \BibitemOpen
  \bibfield  {author} {\bibinfo {author} {\bibfnamefont {Flavio}\ \bibnamefont
  {Romano}}, \bibinfo {author} {\bibfnamefont {Cristiano}\ \bibnamefont
  {De~Michele}}, \bibinfo {author} {\bibfnamefont {Davide}\ \bibnamefont
  {Marenduzzo}}, \ and\ \bibinfo {author} {\bibfnamefont {Eduardo}\
  \bibnamefont {Sanz}},\ }\bibfield  {title} {\enquote {\bibinfo {title} {Monte
  {Carlo} and event-driven dynamics of {Brownian} particles with orientational
  degrees of freedom},}\ }\href {\doibase 10.1063/1.3629452} {\bibfield
  {journal} {\bibinfo  {journal} {J. Chem. Phys.}\ }\textbf {\bibinfo {volume}
  {135}},\ \bibinfo {pages} {124106} (\bibinfo {year} {2011})}\BibitemShut
  {NoStop}%
\bibitem [{\citenamefont {Patti}\ and\ \citenamefont
  {Cuetos}(2012)}]{Patti2012}%
  \BibitemOpen
  \bibfield  {author} {\bibinfo {author} {\bibfnamefont {Alessandro}\
  \bibnamefont {Patti}}\ and\ \bibinfo {author} {\bibfnamefont {Alejandro}\
  \bibnamefont {Cuetos}},\ }\bibfield  {title} {\enquote {\bibinfo {title}
  {Brownian dynamics and dynamic {Monte} {Carlo} simulations of isotropic and
  liquid crystal phases of anisotropic colloidal particles: A comparative
  study},}\ }\href {\doibase 10.1103/PhysRevE.86.011403} {\bibfield  {journal}
  {\bibinfo  {journal} {Phys. Rev. E}\ }\textbf {\bibinfo {volume} {86}},\
  \bibinfo {pages} {011403} (\bibinfo {year} {2012})}\BibitemShut {NoStop}%
\bibitem [{\citenamefont {Cuetos}\ and\ \citenamefont
  {Patti}(2015)}]{Cuetos2015}%
  \BibitemOpen
  \bibfield  {author} {\bibinfo {author} {\bibfnamefont {Alejandro}\
  \bibnamefont {Cuetos}}\ and\ \bibinfo {author} {\bibfnamefont {Alessandro}\
  \bibnamefont {Patti}},\ }\bibfield  {title} {\enquote {\bibinfo {title}
  {Equivalence of {Brownian} dynamics and dynamic {Monte} {Carlo} simulations
  in multicomponent colloidal suspensions},}\ }\href {\doibase
  10.1103/PhysRevE.92.022302} {\bibfield  {journal} {\bibinfo  {journal} {Phys.
  Rev. E}\ }\textbf {\bibinfo {volume} {92}},\ \bibinfo {pages} {022302}
  (\bibinfo {year} {2015})}\BibitemShut {NoStop}%
\bibitem [{\citenamefont {Corbett}\ \emph {et~al.}(2018)\citenamefont
  {Corbett}, \citenamefont {Cuetos}, \citenamefont {Dennison},\ and\
  \citenamefont {Patti}}]{Corbett2018}%
  \BibitemOpen
  \bibfield  {author} {\bibinfo {author} {\bibfnamefont {Daniel}\ \bibnamefont
  {Corbett}}, \bibinfo {author} {\bibfnamefont {Alejandro}\ \bibnamefont
  {Cuetos}}, \bibinfo {author} {\bibfnamefont {Matthew}\ \bibnamefont
  {Dennison}}, \ and\ \bibinfo {author} {\bibfnamefont {Alessandro}\
  \bibnamefont {Patti}},\ }\bibfield  {title} {\enquote {\bibinfo {title}
  {Dynamic {Monte} {Carlo} algorithm for out-of-equilibrium processes in
  colloidal dispersions},}\ }\href {\doibase 10.1039/C8CP02415D} {\bibfield
  {journal} {\bibinfo  {journal} {Phys. Chem. Chem. Phys.}\ }\textbf {\bibinfo
  {volume} {20}},\ \bibinfo {pages} {15118--15127} (\bibinfo {year}
  {2018})}\BibitemShut {NoStop}%
\bibitem [{\citenamefont {Evans}(1993)}]{Evans1993}%
  \BibitemOpen
  \bibfield  {author} {\bibinfo {author} {\bibfnamefont {J.~W.}\ \bibnamefont
  {Evans}},\ }\bibfield  {title} {\enquote {\bibinfo {title} {Random and
  cooperative sequential adsorption},}\ }\href {\doibase
  10.1103/RevModPhys.65.1281} {\bibfield  {journal} {\bibinfo  {journal} {Rev.
  Mod. Phys.}\ }\textbf {\bibinfo {volume} {65}},\ \bibinfo {pages}
  {1281--1329} (\bibinfo {year} {1993})}\BibitemShut {NoStop}%
\bibitem [{\citenamefont {L\"owen}(1994)}]{Loewen1994}%
  \BibitemOpen
  \bibfield  {author} {\bibinfo {author} {\bibfnamefont {Hartmut}\ \bibnamefont
  {L\"owen}},\ }\bibfield  {title} {\enquote {\bibinfo {title} {Brownian
  dynamics of hard spherocylinders},}\ }\href {\doibase
  10.1103/PhysRevE.50.1232} {\bibfield  {journal} {\bibinfo  {journal} {Phys.
  Rev. E}\ }\textbf {\bibinfo {volume} {50}},\ \bibinfo {pages} {1232--1242}
  (\bibinfo {year} {1994})}\BibitemShut {NoStop}%
\bibitem [{\citenamefont {Landau}\ and\ \citenamefont
  {Binder}(2014)}]{Landau2014}%
  \BibitemOpen
  \bibfield  {author} {\bibinfo {author} {\bibfnamefont {David~P.}\
  \bibnamefont {Landau}}\ and\ \bibinfo {author} {\bibfnamefont {Kurt}\
  \bibnamefont {Binder}},\ }\href {\doibase 10.1017/CBO9781139696463} {\emph
  {\bibinfo {title} {A guide to {Monte} {Carlo} simulations in statistical
  physics}}},\ \bibinfo {edition} {4th}\ ed.\ (\bibinfo  {publisher} {Cambridge
  University Press},\ \bibinfo {year} {2014})\BibitemShut {NoStop}%
\bibitem [{\citenamefont {Lebovka}\ \emph
  {et~al.}(2018{\natexlab{a}})\citenamefont {Lebovka}, \citenamefont
  {Tarasevich}, \citenamefont {Vygornitskii}, \citenamefont {Eserkepov},\ and\
  \citenamefont {Akhunzhanov}}]{Lebovka2018PREanisotropy}%
  \BibitemOpen
  \bibfield  {author} {\bibinfo {author} {\bibfnamefont {N.~I.}\ \bibnamefont
  {Lebovka}}, \bibinfo {author} {\bibfnamefont {Yu.~Yu.}\ \bibnamefont
  {Tarasevich}}, \bibinfo {author} {\bibfnamefont {N.~V.}\ \bibnamefont
  {Vygornitskii}}, \bibinfo {author} {\bibfnamefont {A.~V.}\ \bibnamefont
  {Eserkepov}}, \ and\ \bibinfo {author} {\bibfnamefont {R.~K.}\ \bibnamefont
  {Akhunzhanov}},\ }\bibfield  {title} {\enquote {\bibinfo {title} {Anisotropy
  in electrical conductivity of films of aligned intersecting conducting
  rods},}\ }\href {\doibase 10.1103/PhysRevE.98.012104} {\bibfield  {journal}
  {\bibinfo  {journal} {Phys. Rev. E}\ }\textbf {\bibinfo {volume} {98}},\
  \bibinfo {pages} {012104} (\bibinfo {year} {2018}{\natexlab{a}})}\BibitemShut
  {NoStop}%
\bibitem [{\citenamefont {Tarasevich}\ \emph
  {et~al.}(2018{\natexlab{a}})\citenamefont {Tarasevich}, \citenamefont
  {Lebovka}, \citenamefont {Vodolazskaya}, \citenamefont {Eserkepov},
  \citenamefont {Goltseva},\ and\ \citenamefont
  {Chirkova}}]{Tarasevich2018PREb}%
  \BibitemOpen
  \bibfield  {author} {\bibinfo {author} {\bibfnamefont {Yu.~Yu.}\ \bibnamefont
  {Tarasevich}}, \bibinfo {author} {\bibfnamefont {N.~I.}\ \bibnamefont
  {Lebovka}}, \bibinfo {author} {\bibfnamefont {I.~V.}\ \bibnamefont
  {Vodolazskaya}}, \bibinfo {author} {\bibfnamefont {A.~V.}\ \bibnamefont
  {Eserkepov}}, \bibinfo {author} {\bibfnamefont {V.~A.}\ \bibnamefont
  {Goltseva}}, \ and\ \bibinfo {author} {\bibfnamefont {V.~V.}\ \bibnamefont
  {Chirkova}},\ }\bibfield  {title} {\enquote {\bibinfo {title} {Anisotropy in
  electrical conductivity of two-dimensional films containing aligned
  non-intersecting rodlike particles: continuous and lattice models},}\ }\href
  {\doibase 10.1103/PhysRevE.98.012105} {\bibfield  {journal} {\bibinfo
  {journal} {Phys. Rev. E}\ }\textbf {\bibinfo {volume} {98}},\ \bibinfo
  {pages} {012105} (\bibinfo {year} {2018}{\natexlab{a}})}\BibitemShut
  {NoStop}%
\bibitem [{\citenamefont {Tarasevich}\ \emph
  {et~al.}(2018{\natexlab{b}})\citenamefont {Tarasevich}, \citenamefont
  {Vodolazskaya}, \citenamefont {Eserkepov}, \citenamefont {Goltseva},
  \citenamefont {Selin},\ and\ \citenamefont {Lebovka}}]{Tarasevich2018JAP}%
  \BibitemOpen
  \bibfield  {author} {\bibinfo {author} {\bibfnamefont {Yu.~Yu.}\ \bibnamefont
  {Tarasevich}}, \bibinfo {author} {\bibfnamefont {I.~V.}\ \bibnamefont
  {Vodolazskaya}}, \bibinfo {author} {\bibfnamefont {A.~V.}\ \bibnamefont
  {Eserkepov}}, \bibinfo {author} {\bibfnamefont {V.~A.}\ \bibnamefont
  {Goltseva}}, \bibinfo {author} {\bibfnamefont {P.~G.}\ \bibnamefont {Selin}},
  \ and\ \bibinfo {author} {\bibfnamefont {N.~I.}\ \bibnamefont {Lebovka}},\
  }\bibfield  {title} {\enquote {\bibinfo {title} {Simulation of the electrical
  conductivity of two-dimensional films with aligned rod-like conductive
  fillers: Effect of the filler length dispersity},}\ }\href {\doibase
  10.1063/1.5051090} {\bibfield  {journal} {\bibinfo  {journal} {J. Appl.
  Phys.}\ }\textbf {\bibinfo {volume} {124}},\ \bibinfo {pages} {145106}
  (\bibinfo {year} {2018}{\natexlab{b}})}\BibitemShut {NoStop}%
\bibitem [{\citenamefont {Lebovka}\ \emph
  {et~al.}(2018{\natexlab{b}})\citenamefont {Lebovka}, \citenamefont
  {Tarasevich},\ and\ \citenamefont
  {Vygornitskii}}]{Lebovka2018PREverticaldrying}%
  \BibitemOpen
  \bibfield  {author} {\bibinfo {author} {\bibfnamefont {N.~I.}\ \bibnamefont
  {Lebovka}}, \bibinfo {author} {\bibfnamefont {Yu.~Yu.}\ \bibnamefont
  {Tarasevich}}, \ and\ \bibinfo {author} {\bibfnamefont {N.~V.}\ \bibnamefont
  {Vygornitskii}},\ }\bibfield  {title} {\enquote {\bibinfo {title} {Vertical
  drying of a suspension of sticks: {M}onte {C}arlo simulation for continuous
  two-dimensional problem},}\ }\href {\doibase 10.1103/PhysRevE.97.022136}
  {\bibfield  {journal} {\bibinfo  {journal} {Phys. Rev. E}\ }\textbf {\bibinfo
  {volume} {97}},\ \bibinfo {pages} {022136} (\bibinfo {year}
  {2018}{\natexlab{b}})}\BibitemShut {NoStop}%
\bibitem [{\citenamefont {Frank}\ and\ \citenamefont {Lobb}(1988)}]{Frank1988}%
  \BibitemOpen
  \bibfield  {author} {\bibinfo {author} {\bibfnamefont {D.~J.}\ \bibnamefont
  {Frank}}\ and\ \bibinfo {author} {\bibfnamefont {C.~J.}\ \bibnamefont
  {Lobb}},\ }\bibfield  {title} {\enquote {\bibinfo {title} {Highly efficient
  algorithm for percolative transport studies in two dimensions},}\ }\href
  {\doibase 10.1103/PhysRevB.37.302} {\bibfield  {journal} {\bibinfo  {journal}
  {Phys. Rev. B}\ }\textbf {\bibinfo {volume} {37}},\ \bibinfo {pages}
  {302--307} (\bibinfo {year} {1988})}\BibitemShut {NoStop}%
\bibitem [{\citenamefont {Tarasevich}\ \emph {et~al.}(2019)\citenamefont
  {Tarasevich}, \citenamefont {Vodolazskaya}, \citenamefont {Eserkepov},\ and\
  \citenamefont {Akhunzhanov}}]{Tarasevich2019}%
  \BibitemOpen
  \bibfield  {author} {\bibinfo {author} {\bibfnamefont {Y.~Y.}\ \bibnamefont
  {Tarasevich}}, \bibinfo {author} {\bibfnamefont {I.~V.}\ \bibnamefont
  {Vodolazskaya}}, \bibinfo {author} {\bibfnamefont {A.~V.}\ \bibnamefont
  {Eserkepov}}, \ and\ \bibinfo {author} {\bibfnamefont {R.~K.}\ \bibnamefont
  {Akhunzhanov}},\ }\bibfield  {title} {\enquote {\bibinfo {title} {Electrical
  conductance of two-dimensional composites with embedded rodlike fillers: an
  analytical consideration and comparison of two computational approaches},}\
  }\href {\doibase 10.1063/1.5092351} {\bibfield  {journal} {\bibinfo
  {journal} {J. Appl. Phys}\ }\textbf {\bibinfo {volume} {125}},\ \bibinfo
  {pages} {0} (\bibinfo {year} {2019})}\BibitemShut {NoStop}%
\bibitem [{\citenamefont {Lebovka}\ \emph {et~al.}(2016)\citenamefont
  {Lebovka}, \citenamefont {Vygornitskii}, \citenamefont {Gigiberiya},\ and\
  \citenamefont {Tarasevich}}]{Lebovka2016}%
  \BibitemOpen
  \bibfield  {author} {\bibinfo {author} {\bibfnamefont {N.~I.}\ \bibnamefont
  {Lebovka}}, \bibinfo {author} {\bibfnamefont {N.~V.}\ \bibnamefont
  {Vygornitskii}}, \bibinfo {author} {\bibfnamefont {V.~A.}\ \bibnamefont
  {Gigiberiya}}, \ and\ \bibinfo {author} {\bibfnamefont {Yu.~Yu.}\
  \bibnamefont {Tarasevich}},\ }\bibfield  {title} {\enquote {\bibinfo {title}
  {Monte {C}arlo simulation of evaporation driven self-assembly in suspension
  of colloidal rods},}\ }\href {\doibase 10.1103/PhysRevE.94.062803} {\bibfield
   {journal} {\bibinfo  {journal} {Phys. Rev. E}\ }\textbf {\bibinfo {volume}
  {94}},\ \bibinfo {pages} {062803} (\bibinfo {year} {2016})}\BibitemShut
  {NoStop}%
\bibitem [{Note1()}]{Note1}%
  \BibitemOpen
  \bibinfo {note} {\protect \url
  {http://cluster.univ.kiev.ua/eng/}}\BibitemShut {NoStop}%
\end{thebibliography}%

\end{document}